\documentclass[reprint,amsmath,amssymb,aps,pra]{revtex4-2}

\usepackage[utf8]{inputenc}
\usepackage[T1]{fontenc}
\usepackage{graphicx}
\usepackage{dcolumn}
\usepackage{bm}
\usepackage{color}
\usepackage{hyperref}
\usepackage{upgreek,textgreek}
\usepackage{physics}

\usepackage{orcidlink}

\newcommand{\Hop}{\hat{H}} 
\newcommand{\Jop}{\hat{J}} 
\newcommand{\JJop}{\hat{\bm{J}}} 
\newcommand{\ddop}{\hat{\bm{d}}} 

\newcommand{\threej}[6]{\begin{pmatrix}#1 & #2 & #3\\#4 & #5 & #6\end{pmatrix}}

\begin{document}


\title{Rotational properties of two interacting cold polar molecules: linear, symmetric, and asymmetric tops}

\author{Felipe Isaule~\orcidlink{0000-0003-1810-0707}}
\author{Robert Bennett}
\author{J\"org B. G\"otte}
\affiliation{School of Physics and Astronomy, University of Glasgow, Glasgow G12 8QQ,
United Kingdom}


\begin{abstract}

We examine the potential-energy curves and polarization of the dipole moments of two static polar molecules under the influence of an external dc electric field and their anisotropic dipole-dipole interaction. We model the molecules as quantum rigid rotors to take their rotational degrees of freedom into account and consider a selection of linear, symmetric, and asymmetric top molecules. We provide a comprehensive examination of the energy curves and polarization of the dipoles for varying inter-molecular separation and direction of the electric field and find that the properties of the molecules depend strongly on the field's direction at short separations, showing the importance of accounting for molecular rotation. The latter provides insight into the possible effects of accounting for rotational degrees of freedom in molecular dipolar gases.

\end{abstract}

\maketitle

\section{Introduction}

The last two decades have seen unprecedented progress in the realization of cold and controlled molecules~\cite{carr_cold_2009,quemener_ultracold_2012,bohn_cold_2017,langen_quantum_2024}. 
Initially, ultracold molecules were realized  by forming dimers of two cold alkali atoms~\cite{ni_high_2008,takekoshi_ultracold_2014,guo_creation_2016,park_ultracold_2015,molony_creation_2014}. However, rapid progress in cooling techniques has lifted this restriction~\cite{tarbutt_laser_2018,mccarron_laser_2018}, already enabling the trapping of a few ultracold non-alkali diatomic~\cite{norrgard_submillikelvin_2016,truppe_molecules_2017,anderegg_laser_2018,ding_sub-doppler_2020}
and linear triatomic~\cite{kozyryev_sisyphus_2017,augenbraun_laser-cooled_2020,vilas_magneto-optical_2022} 
molecules. In this direction, the symmetric top molecule CaOCH$_3$ has been recently laser-cooled~\cite{mitra_direct_2020} and exciting new developments are expected in the near future, as roadmaps for cooling more complex polyatomic molecules have already been proposed~\cite{isaev_polyatomic_2016,kozyryev_proposal_2016,augenbraun_molecular_2020,xia_production_2022}, even for chiral~\cite{kozyryev_proposal_2016,isaev_towards_2018,augenbraun_molecular_2020} and organic~\cite{ivanov_toward_2020} molecules.

These developments have increased the interest in using ultracold molecules in several applications~\cite{karman_ultracold_2024}, ranging from quantum computing~\cite{demille_quantum_2002,yelin_schemes_2006,wei_entanglement_2011,herrera_infrared-dressed_2014,park_second-scale_2017,hudson_dipolar_2018,ni_dipolar_2018,yu_scalable_2019,gregory_robust_2021,burchesky_rotational_2021,holland_-demand_2023,bao_dipolar_2023,cornish_quantum_2024} 
to probing fundamental physics~\cite{baron_order_2014,andreev_improved_2018,lim_laser_2018,hutzler_polyatomic_2020,mitra_quantum_2022,roussy_improved_2023}. 
Many of these applications rely on the polar nature of the molecules in consideration, enabling further control by electromagnetic fields~\cite{krems_molecules_2019} and the exploitation of the anisotropic dipole-dipole interactions between molecules~\cite{ni_dipolar_2010}. In addition, these features make polar molecules a good option for realizing ultracold dipolar gases~\cite{lahaye_physics_2009,baranov_condensed_2012}, offering new ways to explore novel quantum phases of matter, such as crystalline phases in optical lattices~\cite{kovrizhin_density_2005}. In this direction, degenerate gases of polar diatomic molecules have already been produced~\cite{de_marco_degenerate_2019,valtolina_dipolar_2020}, including the recent realization of a BEC of NaCs molecules~\cite{bigagli_observation_2024}, becoming a rapidly expanding field~\cite{moses_new_2017}. Closely related, spin models with polar molecules in optical lattices have already been produced~\cite{yan_observation_2013,christakis_probing_2023}, as well as polar molecules in optical tweezers arrays~\cite{cairncross_assembly_2021,anderegg_optical_2019,ruttley_enhanced_2024}.

Naturally, molecules offer rich physics due to their complex internal structure~\cite{carr_cold_2009}. Among the internal molecular degrees of freedom, one of the most relevant to consider is internal rotations~\cite{krems_molecules_2019,di_lauro_rotational_2020}. 
Indeed, the control of molecular rotation has attracted significant interest~\cite{koch_quantum_2019}. Rotational modes show a rich response to electromagnetic fields, motivating a plethora of applications, from quantum gates~\cite{wei_entanglement_2011,ni_dipolar_2018,yu_scalable_2019} to chiral discrimination~\cite{tutunnikov_selective_2018}. 
Rich internal structure, however, comes with a cost; experimental molecular control and the theoretical molecular description become more challenging than for atomic systems.

Here we are concerned with a theoretical account of the rotation of interacting polar molecules, which could be probed by examining collisions between molecules. A variety of rotational properties in collisions between ultracold molecules have been studied in the past decade, mostly for diatomic molecules~\cite{micheli_cold_2007,julienne_universal_2011,byrd_controllable_2012,lepers_long-range_2013,wang_tuning_2015,lassabliere_controlling_2018,karman_microwave_2018, xie_optical_2020,walraven_rotational-state_2024}, but also triatomic ones~\cite{augustovicova_ultracold_2019}. In particular, the potential-energy curves of two interacting and rotating polar molecules have been examined in Refs.~\cite{micheli_cold_2007,lepers_long-range_2013,augustovicova_ultracold_2019,walraven_rotational-state_2024}, while the polarization has been studied in Ref.~\cite{lepers_long-range_2013}. Rotational properties of diatomic molecules have also been studied in harmonic traps~\cite{micheli_cold_2007,gorecki_electric_2017,dawid_two_2018,dawid_magnetic_2020} and, very recently, in optical tweezers~\cite{sroczynska_controlling_2022}. Similarly, in many-body lattice scenarios, there have been efforts to include rotational excitations of diatomic molecules in Hubbard-like Hamiltonians~\cite{wall_emergent_2009,herrera_tunable_2011,wall_simulating_2013,wall_realizing_2015}, as well as in the construction of spin models of polar molecules~\cite{wall_realizing_2015,gorshkov_tunable_2011}. However, most works on dipolar gases neglect the effect of rotation, as they either consider magnetic atoms or are motivated by the use of strong electric fields polarizing the molecules in a single rotating state~\cite{lahaye_physics_2009}. Therefore, a good understanding of the role of molecular rotation in inter-molecular interactions is essential for future studies of molecular gases~\cite{micheli_cold_2007,walraven_rotational-state_2024}. In addition, proposed applications of polar molecules to quantum computation rely on the control of rotational excitations~\cite{wei_entanglement_2011,ni_dipolar_2018}, and thus a good understanding of the behavior of interacting rotating molecules is necessary for building robust molecular quantum computing platforms.  

In this exploratory work, we numerically investigate the behavior of two interacting static polar rotating molecules, including linear, symmetric, and asymmetric top molecules. We examine their response when varying their separation and to the introduction of an external dc field. We provide an exhaustive examination of the low-energy spectrum of our model, which provides the potential energy curves of the dipole-dipole interaction of rigid rotors~\cite{micheli_cold_2007}. Additionally, we examine the projection of the electric dipole moment in the direction of the external field for the low energy states, going beyond previous examinations~\cite{lepers_long-range_2013}. The dipole moment projection enables us to quantify the polarization of the molecules by electric fields. In particular, we find that the potential-energy curves and dipole projection depend strongly on the distance between molecules and on the direction of the external dc field. 
Even though our simplified model does not take all effects into account, as we neglect other important degrees of freedom such as hyperfine structure, our findings and techniques used in this work provide a starting point for future realistic studies of polar molecules, such as in many-body scenarios or in the implementation of quantum gates.

This work is organized as follows. In section~\ref{sec:model} we present our theoretical model and the molecules in consideration. In section~\ref{sec:noE} we study two interacting polar molecules in the absence of external fields, focusing on the energy spectrum of our model. Then, in section~\ref{sec:E} we study the interacting molecules under the influence of an external dc field, where we examine both the potential-energy curves and dipole moment projection. Finally, in Sec~\ref{sec:concl} we provide conclusions of our work and an outlook for future studies.

\section{Model}
\label{sec:model}

\begin{figure}[t]
\centering
\includegraphics[width=\columnwidth]{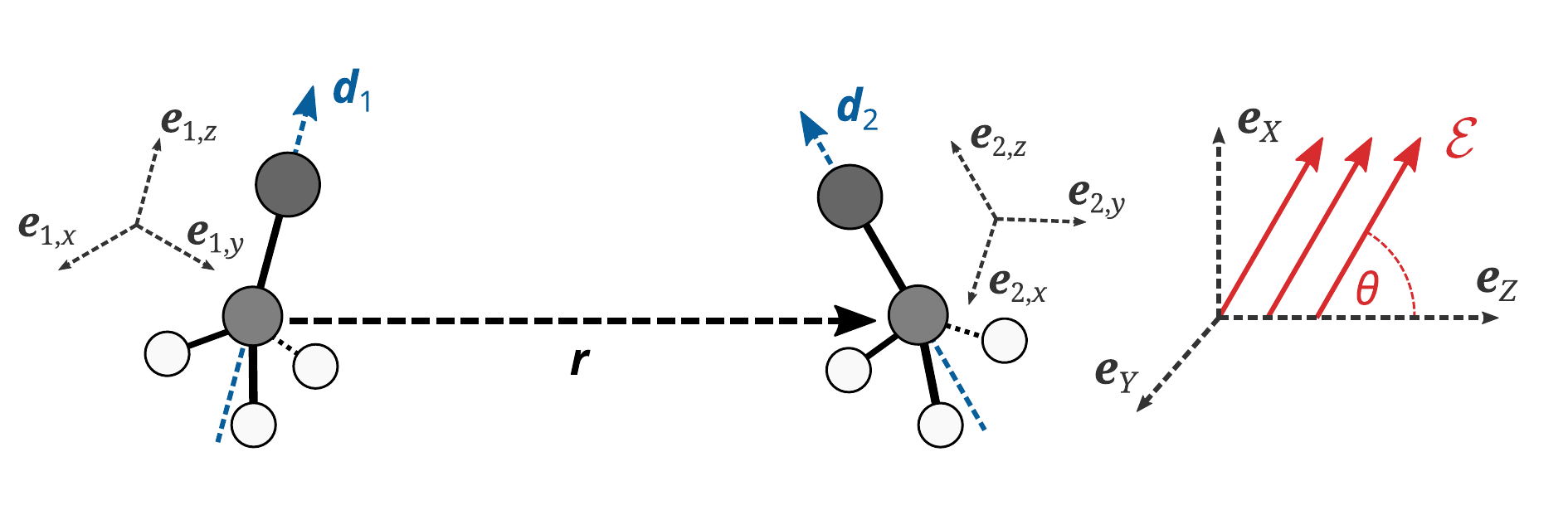}
\caption{Illustration of two rotating polar molecules with permanent dipole moments $\bm{d}_1$ and $\bm{d}_2$ interacting with an external dc field $\mathcal{E}$. The laboratory-fixed frame $(X,Y,Z)$ is chosen so that $\bm{r}=r\bm{e}_Z$. In addition, each molecule has an associated molecule-fixed frame $(x_i,y_i,z_i)$. Note that we illustrate two prolate symmetric tops with $\bm{d}_i=d_i\bm{e}_{i,z}$, but we can employ other choices.}
\label{sec:model;fig:molecules}
\end{figure}

We consider two polar molecules in their ground vibrational state and separated by a distance $r$ as depicted in Fig.~\ref{sec:model;fig:molecules}. Each molecule is modeled as a rigid rotor with a permanent dipole moment $\bm{d}_i$, where $i=1,2$ labels each molecule, giving rise to a dipole-dipole interaction between them. The magnitude of the permanent dipole moment is denoted by $d_i=\norm{\bm{d}_i}$. In addition, the molecules interact with a uniform dc electric field $\bm{\mathcal{E}}$. Note that the molecules are static, and thus $r$ is a fixed distance in our model. However, we study our system for a range of different distances $r$.
In the following, we present our theoretical model and numerical considerations.

\subsection{Hamiltonian}
\label{sec:model;sub:H}

The Hamiltonian describing the system reads as
\begin{equation}
    \Hop = \Hop_\mathrm{rot}+\Hop_\mathrm{dc}+\Hop_\mathrm{dd}\,,
    \label{sec:model;sub:H:eq:H}
\end{equation}
where $\Hop_\mathrm{rot}$ describes the rotation of both molecules, $\Hop_\mathrm{dc}$ describes the interaction with the external electric field, and $\Hop_\mathrm{dd}$ describes the dipole-dipole interaction. Note that the Hamiltonian does not consider the effect of internal spin.

Under the rigid rotor approximation, the rotation of the molecules is described by~\cite{zare_angular_1988}
\begin{equation}
    \Hop_\mathrm{rot}=\sum_{i=1}^2\left(A_i\JJop_{i,a_i}^2+B_i\JJop_{i,b_i}^2+C_i\JJop_{i,c_i}^2\right)\,,
    \label{sec:model;sub:H;eq:Hrot}
\end{equation}
where $A_i \geq B_i \geq C_i$ are the rotational constants of each molecule in their principal axes of inertia $a_i$, $b_i$, and $c_i$, and $\JJop_i=(\Jop_{i,a_i},\Jop_{i,b_i},\Jop_{i,c_i})$ are the angular-momentum operators. The values of the rotational constants define the geometry of the molecule, as detailed in Appendix~\ref{app:single;sub:rot}. We also define the \emph{molecule-fixed} frame as $(x_i, y_i, z_i) = (a_i, b_i, c_i)$ for oblate symmetric tops, whereas as $(x_i, y_i, z_i) = (b_i, c_i, a_i)$ otherwise~\cite{zare_angular_1988}. In the following, we work with two molecules of the same species, and therefore $A=A_1=A_2$, $B=B_1=B_2$, and $C=C_1=C_2$. Note that current experiments with dipolar molecular gases employ identical molecules. In addition, this assumption means that the Hamiltonian~(\ref{sec:model;sub:H:eq:H}) is invariant under the interchange of molecules. This symmetry results in degeneracies~\cite{micheli_cold_2007,lepers_long-range_2013}, which will be examined in the following sections. Nevertheless,  this assumption could easily be relaxed, and could be useful to study collisions between non-identical molecules in the future. \

The interaction between the permanent dipoles and the external field is described by
\begin{equation}
    \Hop_\mathrm{dc}=-\sum_{i=1}^2\ddop_i\cdot\bm{\mathcal{E}}\,,
    \label{sec:model;sub:H;eq:Hdc}
\end{equation}
where $\ddop_i$ is the dipole-moment operator and
\begin{equation}
\bm{\mathcal{E}}=\mathcal{E}( \cos\theta\,\bm{e}_Z+ \sin\theta\,\bm{e}_X)\,,
\end{equation}
is an external dc field. The \emph{laboratory-fixed} frame $(X,Y,Z)$ is chosen such that the vector connecting the two molecules is defined by $\bm{r}=r\bm{e}_Z$, and thus $\theta$ is the angle between $\mathcal{E}$ and the $YZ$-plane (see Fig.~\ref{sec:model;fig:molecules}).

Finally, the dipole-dipole interaction is described by
\begin{equation}
    \Hop_\mathrm{dd}=\frac{(\ddop_1\cdot\ddop_2)-3(\ddop_1\cdot \bm{e}_{\bm{r}})(\ddop_2\cdot\bm{e}_{\bm{r}})}{r^3}\,,
    \label{sec:model;sub:H;eq:Hdd}
\end{equation}
where $\bm{e}_{\bm{r}}= \bm{e}_Z$ in our chosen frame. Note that we work in units of $\epsilon_0=1/4\pi$, where $\epsilon_0$ is the vacuum permittivity. In order to work with the standard rotational basis set, we recast the dipole-dipole interaction in terms of spherical tensors~\cite{zare_angular_1988,man_cartesian_2013}, enabling us to write $\Hop_\mathrm{dd}$ as~\cite{krems_molecules_2019}
\begin{equation}
    \Hop_\mathrm{dd}=-\frac{\sqrt{6}}{r^3}\sum_{p=-2}^{2}(-1)^p C^{(2)}_{-p}(\Omega_r)[\hat{\bm{d}}_1\otimes \hat{\bm{d}}_2]^{(2)}_p\,,
    \label{sec:model;sub:H;eq:Hddsph}
\end{equation}
where $\Omega_r=(\theta_r,\phi_r)$ denotes the polar and azimuthal angles between $\bm{r}$ and the laboratory-fixed frame, $C^{(2)}_{p}$ is an unnormalized spherical harmonic [see Eq.~(\ref{app:single;sub:sph;eq:C})], and
\begin{equation}
        [\bm{d}_1\otimes \bm{d}_2]^{(2)}_p=\sum_{p'=-1}^1\braket{ 1\, p',1\,p-p'}{2\,p} d_{1,p'}d_{2,p-p'}\,,
\end{equation}
is the rank-2 tensor product between the two dipole moments, with $\braket{l_1\, m_1,l_2\,m_2}{l_3\,m_3}$ the Clebsch–Gordan coefficients. The dipoles have been expressed in their rank-1 spherical tensor form $d_0=d_Z$ and $d_{\pm 1}=\mp(d_X\pm i\,d_Y)/\sqrt{2}$. We refer to appendix~\ref{app:single;sub:sph} for complete details on the spherical tensor formalism.

In the chosen laboratory-fixed frame we have that $\Omega_r=(\theta_r,\phi_r)=(0,0)$, and thus only the $p=0$ term in Eq.~(\ref{sec:model;sub:H;eq:Hddsph}) contributes. Therefore, one obtains the simpler expression~\cite{wall_realizing_2015,dawid_two_2018}
\begin{equation}
    \Hop_\mathrm{dd}= -\frac{1}{r^3}\left(2d_{1,0}d_{2,0}+d_{1,-1}d_{2,1}+d_{1,1}d_{2,-1}\right)\,.
    \label{sec:model;sub:H;eq:HddsphZ}
\end{equation}
In this equation, the dipole elements $d_{i,p}$ are expressed in the laboratory-fixed frame and need to be transformed to their known values in the molecule-fixed frame. This is easily done by using spherical tensor transformations [see Eq.~(\ref{app:single;sub:sph;eq:T})].

Finally, it is important to stress that our model is intended to study the effects of the molecular dipole-dipole interaction on the rotational levels in isolation. We have neglected other degrees of freedom with impact on the angular momentum, most importantly hyperfine structure~\cite{krems_molecules_2019,blackmore_controlling_2020}, but also vibration in polyatomic molecules~\cite{jadbabaie_characterizing_2023}. Our model also neglects other inter-molecular interactions, such as van der Waals contributions~\cite{lepers_long-range_2013}. Nevertheless, our simplified model enables us to isolate the behavior of rotation to be able to model realistic scenarios in future work.

\subsection{Symmetric top basis and diagonalization}

To diagonalize the Hamiltonian, we work in terms of a basis set
\begin{equation}
    \ket{\alpha}  \equiv \ket{j_1\,k_1\,m_1,j_2\,k_2\,m_2}\,,
\end{equation}
 where $\ket{j_i\,k_i\,m_i}$ are the usual symmetric top wavefunctions for each molecule~\cite{zare_angular_1988} which characterize their rotational state. Indeed, the square of the molecule's angular momentum $\JJop_i^2$ has the usual quantization $j_i(j_i+1)$ with $j_i=0,1,...$, whereas
the projection on the laboratory-fixed axis $J_Z$ is quantized by $-j_i\leq m_i \leq j_i$. The additional quantum number $k_i$ denotes the quantization of the projection on the molecule-fixed axis $J_{z_i}$. For linear molecules, it can be discarded ($k=0$), while for non-linear molecules it takes values $-j_i\leq k_i \leq j_i$. A detailed presentation of this basis is provided in Appendix~\ref{app:single}.

 To diagonalize the problem numerically we construct a basis set with Fock states up to a chosen cutoff $j_{\mathrm{max}}$ for both $j_i$. The Hamiltonian matrix is constructed in this truncated subspace, while the diagonalization is afterward performed with standard numerical routines for sparse matrices~\cite{lehoucq_arpack_1998}. In this work, we employ a cutoff of $j_{\mathrm{max}}=7$, which enables us to safely study the lower part of the energy spectrum of our Hamiltonian~\cite{dawid_magnetic_2020}. As mentioned, the spectrum enables us to compute the potential-energy curves associated with the dipole-dipole interaction between two polar molecules modeled as rigid rotors~\cite{lepers_long-range_2013}. The explicit expressions for the matrix elements are provided in Appendix~\ref{app:matrix}. Additionally, an examination of the convergence of the calculations as a function of $j_{\mathrm{max}}$ is presented in Appendix~\ref{app:convergence}. 

\subsection{Molecules under consideration}
\label{sec:model;sub:param}

Even though this is an exploratory theoretical work, we choose known rotational constants of selected molecules. This work considers three types of molecules. First, linear molecules, such as CaF and RbCs, with permanent dipole moments $\bm{d}_i=d\bm{e}_{z_i}$. Because we scale the results in terms of the parameters of the problem, our results are independent of the specific linear molecule in consideration. However, for non-linear top molecules, the results depend on the ratios between the rotational constants, and thus one needs to choose a combination of these~\cite{zare_angular_1988}. To consider realistic rotational constants, this work considers fluoroform (CHF$_3$), an oblate symmetric top, and 1,2-propanediol (CH$_3$CHOHCH$_2$OH), a near-prolate chiral asymmetric top molecule.  Both CHF$_3$ and 1,2-propanediol are not among the current candidates for cooling to ultracold temperatures in the $\mu$K regime and below.  However, their properties are well-known and the values of their rotational constants are convenient to visualize their low-energy properties in this exploratory theoretical study. Nevertheless, both molecules can still be cooled to the $\sim$K regime~\cite{strebel_improved_2010, patterson_cooling_2012}.
\begin{table}[t]
\caption{Molecular properties of CHF$_3$~\cite{schoolery_molecular_1951,meerts_avoidedcrossing_1981} and 1,2-propanediol~\cite{lovas_microwave_2009}. The rotational constants ($A$, $B$, $C$) are given in GHz, whereas the dipole moments ($d_a$, $d_b$, $d_c$) are given in Debye.
\label{sec:model;sub:param;table:param}}
\begin{ruledtabular}
\begin{tabular}{lcccccc}
Molecule & $A$ & $B$ & $C$ & $d_a$ & $d_b$ & $d_c$\\
\colrule
CHF$_3$ & 10.348 & 10.348 & 5.6734 & 0 & 0 & 1.645\\
1,2-propanediol & 8.57205 & 3.640 & 2.790 & 1.2 & 1.9 & $\pm$ 0.36
\end{tabular}
\end{ruledtabular}
\end{table}
\begin{figure*}[t]
\centering
\includegraphics[width=\textwidth]{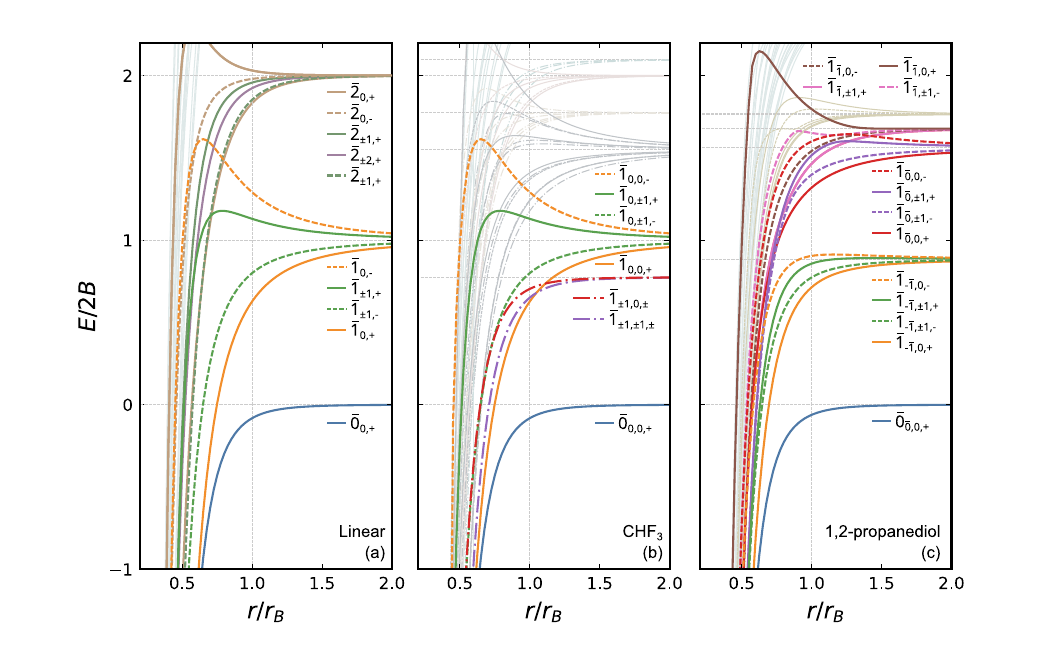}
\caption{Potential-energy curves as a function of $r$ for two linear (a), CHF$_3$ (b), and 1,2-propanediol (c) molecules in the absence of external fields. Solid lines correspond to symmetric solutions, dashed lines correspond to antisymmetric solutions, and dash-dotted states correspond to degenerate solutions of symmetric and antisymmetric states. The thin horizontal lines correspond to the known energies of two independent rotating molecules [see Appendix~\ref{app:single}], while the vertical lines indicate $r=r_B$. The labeling convention is explained in the main text. Additionally, the results shown in panel (a) were reported previously in Ref.~\cite{micheli_cold_2007}.}
\label{sec:noE;fig:spectrum}
\end{figure*}

We provide the rotational constants and components of the dipole moment in table~\ref{sec:model;sub:param;table:param}.
Note that while the opposite enantiomers ($R$ and $S$) of 1,2-propanediol have components $d_c$ with opposite signs ($d^{(R)}_c=-d^{(S)}_c$), the choice of enantiomers does not affect the calculations because our model does not consider any enantio-discriminatory element.

\section{Zero external field}
\label{sec:noE}

We start by examining the behavior of the potential-energy curves of two interacting polar molecules in the absence of external fields.  This enables us to quantify the impact of the dipole-dipole interaction on the rotational states. Note that related examinations for linear molecules have been performed in the past~\cite{micheli_cold_2007,lepers_long-range_2013}, while a symmetric top has been examined a few years ago in Ref.~\cite{augustovicova_ultracold_2019}. However, here we also extend these studies to asymmetric molecules.

In Fig.~\ref{sec:noE;fig:spectrum} we show curves of the potential energy, $E$, as a function of the distance between molecules $r$ for the three types of molecules considered in this work. The energies are scaled in terms of the rotational constant $B$ of each molecule and their separation in terms of the characteristic distance $r_B=(d^2/B)^{1/3}$~\cite{micheli_cold_2007,wall_emergent_2009}. This characteristic distance separates the regime where the dipolar interaction strongly mixes different rotation states ($r\ll r_B$) to that where the dipolar interaction does not significantly affect the non-interacting rotational states ($r\gg r_B$). It is usually of the order of tens of nanometres~\cite{wall_emergent_2009}, much smaller than the average separation that is achieved in experiments~\cite{ni_high_2008}. However, on one side, this regime enables us to better visualize the effect of molecular rotation, as also recently considered in Ref.~\cite{walraven_rotational-state_2024} to examine inter-molecular interactions. Furthermore, this regime is nevertheless important to construct effective dipole-dipole interactions for dipolar gases~\cite{micheli_cold_2007}. Finally, also note that this figure (and all the subsequent ones) only highlights the lowest-energy states to aid readability.

Firstly, at large distances $r\gg r_B$ the molecules decouple, and thus we recover the known potential-energy curves for two non-interacting rotating molecules (as examined in Appendix~\ref{app:single;sub:rot}), as expected. Indeed, for $r\gg r_B$, the ground states in Fig.~\ref{sec:noE;fig:spectrum} simply correspond to two molecules in their $j_i=0$ state, the subsequent excited states correspond to one molecule in its first excited state with the other in its ground state [$(j_1,j_2)=(1,0)$ or $(j_1,j_2)=(0,1)$], and so on. These known energy values are shown as the thin horizontal lines.

For shorter distances $r\lesssim r_b$, as shown in Ref.~\cite{micheli_cold_2007} for linear molecules,  we observe that the dipole-dipole interaction plays a significant role, changing the energy spectrum and breaking degeneracies. Because the system is invariant under the exchange of molecules, we can employ this symmetry to characterize the energy levels~\cite{lepers_long-range_2013}. While the dipole-dipole interaction mixes states with different $j=j_1+j_2$, both $m=m_1+m_2$ and $k=k_1+k_2$ are conserved. Moreover, the solutions can either be symmetric ($+$) or antisymmetric ($-$), which are denoted via $P=\pm$. Thus the full labelling system is: $\bar{j}_{m,P}$ for linear molecules, $\bar{j}_{k,m,P}$ for symmetric molecules, and $\bar{j}_{\bar{\tau},m,P}$ for asymmetric molecules. Here $\bar{j}$ corresponds to the value of $j$ in the limit $r\to\infty$. Note that for $j>1$, $j$ can label different combinations of $j_1$ and $j_2$, such as $(j_1,j_1)=(1,1)$ and $(j_1,j_2)=(2,0)$ for $j=2$. Similarly, $\bar{\tau}$ corresponds to the value of the pseudo-quantum number $\tau=\tau_1+\tau_2$ which labels the states of asymmetric tops (see Appendix~\ref{app:single;sub:rot}). We note that here and in the following we use overbars to denote labels that do not correspond to good quantum numbers.

First, and as reported previously in Refs.~\cite{micheli_cold_2007,lepers_long-range_2013}, for linear tops (a) we observe that as $r$ decreases, the energy levels separate into symmetric (solid lines) and antisymmetric (dashed lines) solutions with a constant $\abs{m}$. Therefore, the levels with $m=0$ are not degenerate, whereas the $|m|>0$ levels are twice degenerate. The gaps between the energy levels become more prominent around $r\approx r_B$, at which point some energy levels increase with decreasing $r$, while others decrease. However, all the energy levels diverge to $E_{\bar{j},m,P}\to -\infty$ deep in the molecular region $r\ll r_B$~\cite{micheli_cold_2007} as a consequence of the diverging nature of the dipole-dipole interaction for $r\to 0$. More importantly, the energy gap between states with different $|m|$ gives rise to spin-exchange interactions, as used in spin models of polar molecules~\cite{gorshkov_tunable_2011,moses_new_2017}.

The potential-energy curves of symmetric tops (b) have a similar structure to that of linear tops. The curves show symmetric and antisymmetric solutions, and solutions with constant $\abs{m}$ and $|k|$. The solutions for $k=0$ simply correspond to the potential-energy curves of linear tops (see blue, orange, and green lines) because symmetric tops do not mix states with different $k$. On the other hand, the solutions for $k\neq 0$ add additional states between those for $k=0$. Note that because we examine an oblate top, the states with $k\neq 0$ have less energy than that for $k=0$. In contrast, states with $k\neq 0$ in prolate tops increase the energy (see Eq.~\ref{app:single;sub:rot;Hrot_pl}).   Interestingly, the states with $k\neq 0$ have degenerate symmetric and antisymmetric solutions (dash-dotted lines). Therefore, the states $\bar{1}_{\pm 1, \pm 1}$ (purple lines) are eightfold degenerate, while the states $\bar{1}_{\pm 1, 0}$ (red lines) are fourfold degenerate. As mentioned, Ref.~\cite{augustovicova_ultracold_2019} studied the collision of two symmetric tops, resulting in very similar energy curves. However, that work also considers additional degrees of freedom, resulting in additional states and symmetries.

Finally, the potential-energy curves of asymmetric tops (c) show similar features to the previous spectra, but with much more energy states. Because asymmetric tops mix states with different $k$, the solutions are only labeled by the pseudo-quantum number $\tau$ (see Appendix~\ref{app:single;sub:rot} for details), and no state coincides with states of linear molecules. Indeed, despite being similar, even the ground state (blue line) of two asymmetric tops is slightly different from that of linear molecules. In addition, asymmetric tops show fewer degenerate states than symmetric tops, as only states with $m\neq 0$ are twice degenerate, with no degenerate symmetric and antisymmetric states.

Because the potential-energy curves of two interacting asymmetric tops depend on the rotational constants, the control of the separation between interacting molecules, for example with optical tweezers, could be employed to experimentally measure the moments of inertia, which are generally difficult to characterize. Moreover, the use of external fields, as in the next section, provides additional parameters to control for molecular measurements. We provide additional details about the dependence of the energy spectrum of asymmetric molecules on their geometry in Appendix~\ref{app:asym}.

It is important to stress that the discussed degeneracies and symmetric and antisymmetric solutions are due to the consideration of identical molecules. Consideration of non-identical would result in different spectra with no degenerate states because, for example, having molecules in states $(j_1,j_2)=(1,0)$  results in different energies to those in states $(j_1,j_2)=(0,1)$. 

\section{Finite external field}
\label{sec:E}

We now turn our attention to the problem of two interacting polar molecules in the presence of an external dc electric field $\bm{\mathcal{E}}$. We continue our examination of the potential-energy curves, and we also examine the total expectation value of the projection of the dipole moments on a laboratory axis $L=X,Y,Z$,
\begin{equation}
    \expval{d_L} = \expval{d_{1,L}+d_{2,L}}/2\,.
\end{equation}
The expectation values are calculated from the obtained eigenvectors (for more details see Appendix~\ref{app:single;sub:dc}). The calculation of the average dipole moments enables us to examine the polarization of the molecules for different distances and choices of the external field. In addition, we have found that the difference between the dipole moments
\begin{equation}
    \langle \Delta d_L\rangle=|\langle d_{1,L}-d_{2,L}\rangle|\,,
    \label{sec:E;eq:DdL}
\end{equation} 
is zero in all our calculations ($\langle \Delta d_L\rangle=0$). Therefore, the molecules are never anti-oriented, which can be expected from the application of the external electric field.

\begin{figure*}[t]
\centering
\includegraphics[width=\textwidth]{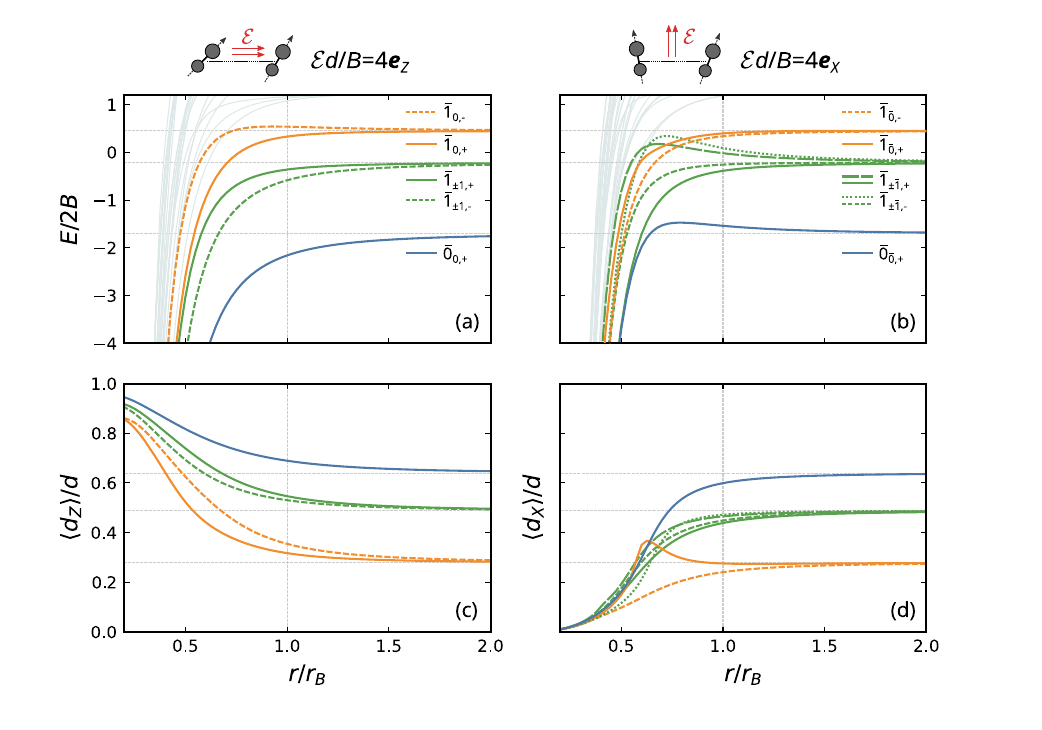}
\caption{Potential-energy curves (a,b) and average dipole moment projection on the $\bm{e}_Z$ (c) and $\bm{e}_X$ (d) axes of two linear molecules as a function of the distance between molecules $r$. The left panels (a,c) consider an external dc field $\bm{\mathcal{E}}d/B=4\bm{e}_Z$, whereas the right panels (b,d) consider an external dc field $\bm{\mathcal{E}}d/B=4\bm{e}_X$. Lines with the same color and line type in the upper and bottom panels indicate the same eigenstate.  The thin horizontal lines correspond to the known energies of two independent rotating molecules interacting with a dc field of magnitude $\mathcal{E}d/B=4$, while the vertical lines indicate $r=r_B$. The labeling convention is explained in the main text.}
\label{sec:E;sub:linear;fig:r}
\end{figure*}

In this section we employ an electric field with a strength $\mathcal{E}d/B=4$. This choice is an intermediate strength large enough to polarize the molecules, but also weak enough to show the competition between rotation, the dipole-dipole interactions, and electric field~\cite{lepers_long-range_2013}. Note that many experimental applications employ strong electric fields to better control the molecules. Nevertheless, similar results and conclusions are obtained with other values of $\mathcal{E}$. A further examination of the dependence on the strength $\mathcal{E}$ is also provided in Appendix~\ref{app:E}.

In the following, we separate our discussion again into linear, symmetric, and asymmetric molecules.

\subsection{Linear tops}
\label{sec:E;sub:linear}

We start by examining the behavior of linear molecules. Similar studies of diatomic molecules in the presence of external electric fields have been reported in Refs.~\cite{micheli_cold_2007,lepers_long-range_2013}, including an examination of the orientation of the ground state in the latter. We consider molecules interacting with a fixed external field as a function of their separation, expanding on the results shown in Fig.~\ref{sec:noE;fig:spectrum}. In Fig.~\ref{sec:E;sub:linear;fig:r} we report the potential-energy curves and average projection of the dipole moment on the laboratory-frame axes as a function of the distance. We consider an external field in the $\bm{e}_Z$ direction in the left panels and a field in the $\bm{e}_X$ direction in the right panels. Because the molecules develop a finite dipole projection only in the direction of the dc field, we only show $\expval{d_Z}$ and $\expval{d_X}$ for the fields in the $\bm{e}_Z$ and $\bm{e}_X$ directions in panels (c) and (d), respectively. We also only show the dipole projections of the highlighted states in the potential-energy curves. Note that all the results converge to the known single molecule limits (thin horizontal lines) for $r\gg r_B$, as expected.

Firstly, note that the symmetries of the Hamiltonian undergo an important change as the direction of the electric field is varied. While $m$ is conserved for fields in the $\bm{e}_Z$ direction, this symmetry is broken if the dc field has a component in any other direction. We can understand this by noting that the dipole-dipole interaction conserves the projection of the angular momentum in the direction of $\bm{r}$, which corresponds to $\bm{e}_Z$ in our case. Therefore, the dipole-dipole interaction conserves $m$, as discussed in the previous section. On the other hand, the electric field conserves the projection of the angular momentum in the direction of the field. Therefore, $m$ is only conserved if the electric field is in the $\bm{e}_Z$ direction. In any other case, there is a competition between the electric field and dipole-dipole interaction which breaks this symmetry.  Due to this, we label the eigenstates in the left panels as $\bar{j}_{m,P}$ as in section~\ref{sec:noE}, whereas we label the states in the right panels as $\bar{j}_{\bar{m},P}$ where $\bar{m}$ gives the value of $m$ in the limit $r\gg r_B$. However, around the molecular region ($r\approx r_B$) the value of $\bar{m}$ is only a label, rather than a good quantum number.

Another interesting consequence of the direction of the field is the breaking of degeneracies, as also reported in Ref.~\cite{micheli_cold_2007}. While for fields in the $\bm{e}_Z$ direction, the two pairs of states with equal $\abs{m}$ are degenerate as with no external field (see the two green lines in the left panels of Fig.~\ref{sec:E;sub:linear;fig:r}), this degeneracy is broken if the field has a component in any other direction (see the four green lines in the right panels). 

Concerning the dipole projections, in all cases, $\expval{d_L}$ converges to the expected single molecule solution for $r\gg r_B$. However, $\expval{d_L}$ behaves significantly differently at short distances depending on the direction of the fields. While $\expval{d_Z}$ increases at short distances for fields in the $\bm{e}_Z$ direction (c), $\expval{d_Z}$ instead decreases at short distances for fields in the $\bm{e}_X$ direction (d), vanishing for $r\to 0$. To interpret this, we recall that the dipole-dipole interaction becomes maximally attractive for dipoles polarized in the direction of $\bm{r}$, whereas it becomes maximally repulsive for dipoles polarized orthogonal to $\bm{r}$.  Therefore, considering that the external field tries to polarize the dipoles in the direction of $\bm{\mathcal{E}}$, the dipole-dipole interaction enhances the polarization at short distances if the field is in the $\bm{e}_Z$ direction. In contrast, for fields in the $\bm{e}_X$ direction, it is not favorable to have dipoles polarized in the $\bm{e}_X$ direction at short distances due to the repulsive nature of the dipole-dipole interaction. This drives $\expval{d_X}$ to vanish for $r\to 0$. Because $\expval{\Delta d_X}=0$ [Eq.~(\ref{sec:E;eq:DdL})], $\expval{d_X}\to 0$ means that the molecules lose their orientation in the $X$-direction. This effect could be relevant to take into account in molecular dipolar gases under weak electric fields. Note that Ref.~\cite{lepers_long-range_2013} reported a similar behavior of the ground-state dipole projections for short inter-molecular separations.

\begin{figure}[t]
\centering
\includegraphics[width=\columnwidth]{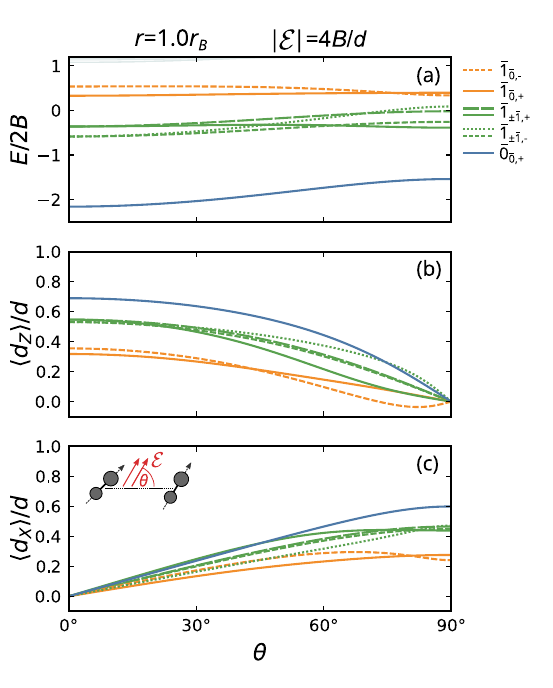}
\caption{Potential-energy curves (a) and average dipole moment projections $\expval{d_Z}$ (b) and  $\expval{d_X}$ (c) for two linear molecules as a function of the angle $\theta$ between $\bm{\mathcal{E}}$ and the YZ-plane. We consider a distance $r=r_B$ and a field of magnitude $\mathcal{E}d/B=4$. Lines with the same color and line type in the upper and bottom panels indicate the same eigenstate. The labeling convention is explained in the main text.}
\label{sec:E;sub:linear;fig:theta}
\end{figure}

Finally, to further examine the behavior of the molecules interacting with dc fields in different directions, we show the potential-energy curves and dipole projections as a function of $\theta$ in Fig.~\ref{sec:E;sub:linear;fig:theta}, going beyond what has been shown in the past~\cite{lepers_long-range_2013}. We choose a distance $r=r_B$ between molecules. The figure connects the field in the $\bm{e}_Z$ direction ($\theta=0^\circ$) to the field in the $\bm{e}_X$ direction ($\theta=90^\circ$). The potential-energy curves show how the presence of components of $\bm{\mathcal{E}}$ orthogonal to $\bm{e}_Z$ breaks the degeneracy of states with equal $\abs{m}$. Indeed, the two green lines at $\theta=0$ smoothly separate into four lines as $\theta$ increases. In addition, the potential-energy curves show how the eigenstates are connected between the two limits, which is the reason for the particular color scheme used in Fig.~\ref{sec:E;sub:linear;fig:r}. Moreover, the figure shows clearly that the direction of the dc field also affects the energies of the levels themselves. Indeed, even the ground state energy depends on $\theta$.

\begin{figure*}[t]
\centering
\includegraphics[width=\textwidth]{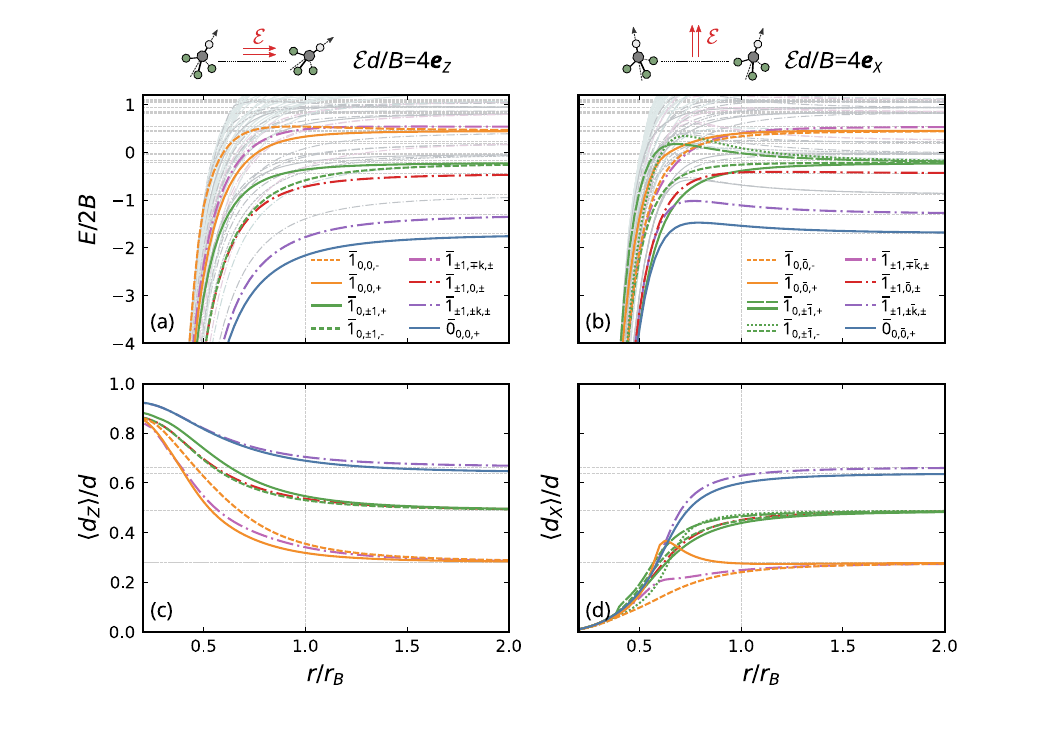}
\caption{Potential-energy curves (a,b) and average dipole moment projection on the $\bm{e}_Z$ (c) and $\bm{e}_X$ (d) axes of two CHF$_3$ molecules as a function of the distance between molecules $r$. The left panels (a,c) consider an external dc field $\bm{\mathcal{E}}d/B=4\bm{e}_Z$, whereas the right panels (b,d) consider an external dc field $\bm{\mathcal{E}}d/B=4\bm{e}_X$. Lines with the same color and line type in the upper and bottom panels indicate the same eigenstate.  The thin horizontal lines correspond to the known energies of two independent rotating molecules interacting with a dc field of magnitude $\mathcal{E}d/B=4$, while the vertical lines indicate $r=r_B$. The labeling convention is explained in the main text.}
\label{sec:E;sub:CHF3;fig:r}
\end{figure*}

 Fig.~\ref{sec:E;sub:linear;fig:theta} also shows how the dipole moment projections depend on the direction of the external field, with $\expval{d_Z}$ developing an increasing finite value as $\theta\to 0$, whereas $\expval{d_X}$ instead develops a finite value as $\theta\to 90^\circ$. Also note that $\langle d_X \rangle|_{\theta=90^\circ}<\langle d_Z\rangle|_{\theta=0^\circ}$ for the same eigenstate due to the enhancement of the dipole projection for fields in the direction of $\bm{r}$, as discussed previously.

\subsection{Symmetric tops}
\label{sec:E;sub:CHF3}

We now extend our previous examination to the oblate symmetric top. The potential-energy curves and dipole moment projections as a function of the distance between CHF$_3$ molecules are shown in Fig.~\ref{sec:E;sub:CHF3;fig:r}, considering again electric fields in the $\bm{e}_Z$ and $\bm{e}_X$ directions. Naturally, the features observed with linear molecules are largely carried over to more complex molecules. Indeed, $m=m_1+m_2$ is only conserved if the dc field is in the direction of $\bm{r}$. However, the projection of the angular momentum onto the molecular axis, $k$, remains conserved in all cases. This can be expected, as the molecule-fixed axes simply move alongside the molecules and do not depend on the direction of the interactions. Therefore, we label the states as $\bar{j}_{k,m,P}$ for fields in the $\bm{e}_Z$ direction, and $\bar{j}_{k,\bar{m},P}$ otherwise, where $\bar{m}$ gives the value of $m$ for $r\gg r_B$.

The dipole projections also behave as they do with linear molecules, as expected. The projection $\expval{d_Z}$ increases at short distances for fields in the $\bm{e}_Z$ direction, whereas $\expval{d_Z}$ decreases for fields in the $\bm{e}_X$ direction. In addition, the states with $k=0$ are identical to those of linear molecules, as there is no mixing of states with different $k$. Nevertheless, the states with $k\neq 0$ fill the energy spectrum with additional states, adding additional energy curves. Again, we only highlight in color the states with $\bar{j}=0,1$.

\begin{figure}[t]
\centering
\includegraphics[width=\columnwidth]{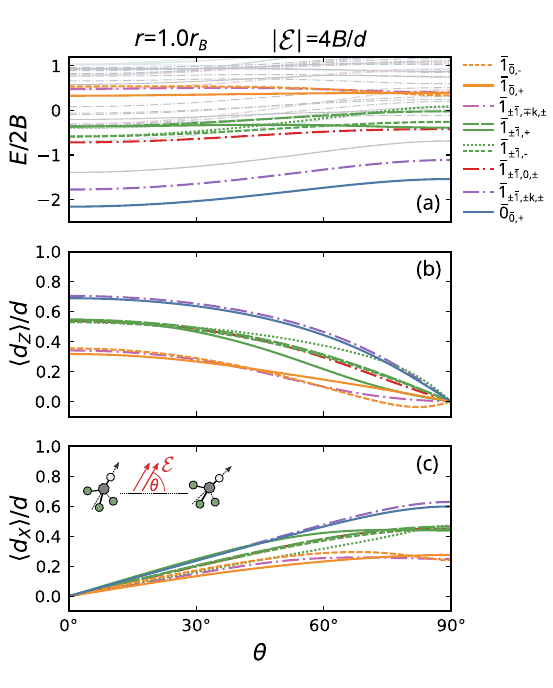}
\caption{Potential-energy curves (a) and average dipole moment projections $\expval{d_Z}$ (b) and  $\expval{d_X}$ (c) for two CHF$_3$ molecules as a function of the angle $\theta$ between $\bm{\mathcal{E}}$ and the YZ-plane. We consider a distance $r=r_B$ and a field of magnitude $\mathcal{E}d/B=4$. Lines with the same color and line type in the upper and bottom panels indicate the same eigenstate. The labeling convention is explained in the main text.}
\label{sec:E;sub:CHF3;fig:theta}
\end{figure}

As seen with no external fields, the states with $k\neq 0$ generally have degenerate symmetric and asymmetric solutions. This degeneracy is not necessarily broken by the introduction of external fields. Indeed, the highlighted excited states (red and purple lines) maintain this degeneracy. However, we have observed that other states can have this degeneracy broken.  The external field does separate states with $k=\pm m$ and $k=\mp m$ (compare purple lines in Fig.~\ref{sec:E;sub:CHF3;fig:theta} with that in Fig.~\ref{sec:noE;fig:spectrum}b). This is simply induced by the Stark effect, as shown for single molecules in Fig.~\ref{app:single;sub:dc;fig:E_dZ}b.

Finally, the potential-energy curves and dipole projections as a function of $\theta$ are shown in Fig.~\ref{sec:E;sub:CHF3;fig:theta}. All the eigenstates show a smooth crossover between the $\theta=0$ and $\theta=90^\circ$ limits, with $\expval{d_X}$ vanishing for $\theta\to 0$, and $\expval{d_Z}$ vanishing for $\theta\to 90^\circ$. The states with $k\neq 0$ do not show any unexpected behavior as $\theta$ changes.

To conclude this section, it is important to note that the added complexity brought by symmetric top molecules compared to linear ones makes them less viable candidates for realizing ultracold dipolar gases. However, as shown by the potential-energy curves, the additional states with finite $k$ open new channels that could be controlled. In this direction, proposals for using such states for constructing quantum gates with interacting symmetric top molecules have been discussed in the past~\cite{yu_scalable_2019}.

\subsection{Asymmetric tops}
\label{sec:E;sub:12prop}

\begin{figure*}[t]
\centering
\includegraphics[width=\textwidth]{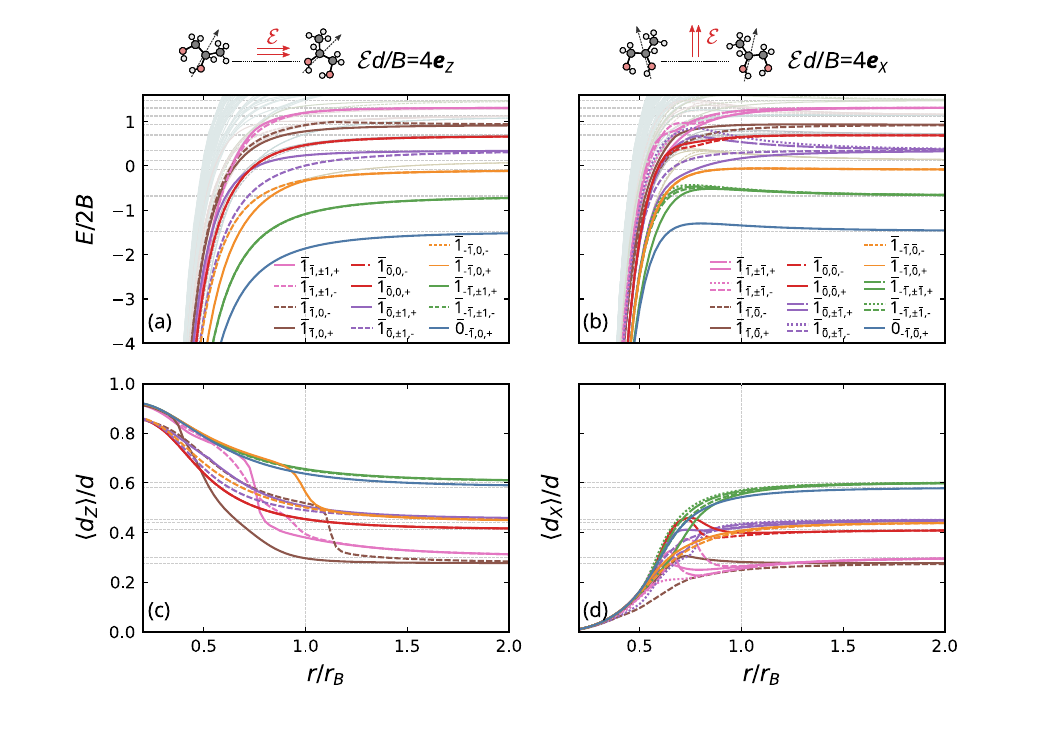}
\caption{Potential-energy curves (a,b) and average dipole moment projection on the $\bm{e}_Z$ (c) and $\bm{e}_X$ (d) axes of two 1,2-propanediol molecules as a function of the distance between molecules $r$. The left panels (a,c) consider an external dc field $\bm{\mathcal{E}}d/B=4\bm{e}_Z$, whereas the right panels (b,d) consider an external dc field $\bm{\mathcal{E}}d/B=4\bm{e}_X$. Lines with the same color and line type in the upper and bottom panels indicate the same eigenstate.  The thin horizontal lines correspond to the known energies of two independent rotating molecules interacting with a dc field of magnitude $\mathcal{E}d/B=4$, while the vertical lines indicate $r=r_B$. The labeling convention is explained in the main text.}
\label{sec:E;sub:12prop;fig:r}
\end{figure*}

We finish our examination of molecules interacting with external dc fields by considering asymmetric tops. The potential-energy curves and dipole projections of two 1,2-propanediol molecules as a function of the separation between them are shown in Fig.~\ref{sec:E;sub:12prop;fig:r}. The figure considers the same fields as in the previous figures. The overall behavior of the potential-energy curves and dipole projections is similar to that of symmetric and linear tops, with the states converging to the known single molecule solutions for $r\gg r_B$ and with $m$ being a good quantum number only for fields in the $\bm{e}_Z$ direction. Similarly, $\expval{d_Z}$ is enhanced at short distances for fields in the $\bm{e}_Z$ direction, whereas $\expval{d_X}$ vanishes at short distances for fields in the $\bm{e}_X$ direction. Therefore, we label the states as $\bar{j}_{\bar{\tau},\bar{m},P}$, extending our previous conventions. Nevertheless, asymmetric tops show a more complex behavior with additional states, which can be expected from the mixing of states with different values of $k$.

As with no external field, and in contrast to symmetric tops, asymmetric molecules do not show degenerate symmetric and asymmetric states and instead, they are noticeably distinguishable for $r\lesssim r_B$. Some states also show prominent jumps in the dipole projections (see for example solid orange and pink lines in panel c) due to avoided crossings in the spectra~\cite{yu_scalable_2019}.  As the molecules move closer (for $r\lesssim r_B$) and many levels have similar energies, there is a strong mixing of Fock states with different $j$ due to the dipole-dipole interaction and of different $k$ due to the rotation of the asymmetric molecules. This mixing produces noticeable variations in the eigenvectors, and thus in the dipole projections. We stress again that the numbers we have associated with each state are only meaningful at $r\gg r_B$, and at shorter distances, they only act as labels of each quantum state. 

\begin{figure}[t]
\centering
\includegraphics[width=\columnwidth]{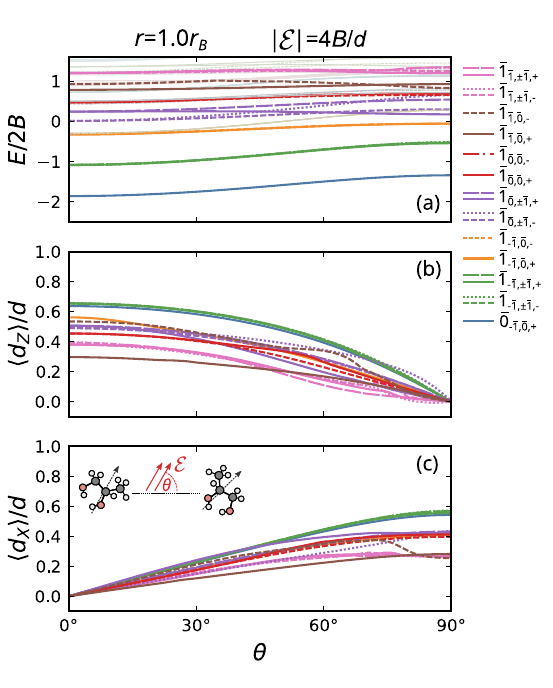}
\caption{Potential-energy curves (a) and average dipole moment projections $\expval{d_Z}$ (b) and  $\expval{d_X}$ (c) for two 1,2-propanediol molecules as a function of the angle $\theta$ between $\bm{\mathcal{E}}$ and the YZ-plane. We consider a distance $r=r_B$ and a field of magnitude $\mathcal{E}d/B=4$. Lines with the same color and line type in the upper and bottom panels indicate the same eigenstate. The labeling convention is explained in the main text.}
\label{sec:E;12prop;fig:theta}
\end{figure}

Finally, the potential-energy curves and dipole projections of two 1,2-propanediol molecules as a function of $\theta$ are shown in Fig.~\ref{sec:E;12prop;fig:theta}. The molecules show the expected behavior, with a smooth crossover between fields in the $\bm{e}_Z$ and $\bm{e}_X$ limits. The additional complexity of asymmetric tops may maybe more noticeable, with a less clear difference in the dipole projections between states. In addition, avoided crossings with jumps in $\langle d \rangle$ also occur as $\theta$ changes.

The mentioned additional complexity of asymmetric tops and high level of near degeneracy makes them less suitable candidates for molecular control. In particular, symmetric tops are probably much better options for quantum computing applications due to having better-defined properties~\cite{yu_scalable_2019}. Nevertheless, a good understanding of the rotational properties of asymmetric molecules can be important for further control of cold chiral molecules, such as for enantio-discrimination~\cite{tutunnikov_selective_2018} or rotational spectroscopy~\cite{cameron_chiral_2016}. In this direction, a more comprehensive study of the states of asymmetric molecules is needed, which is left for future work.

\section{Conclusions}
\label{sec:concl}

We have investigated the behavior of two rotating polar molecules interacting via the dipole-dipole interaction and with an external dc electric field. We presented a comprehensive examination of the potential-energy curves and polarization of the dipoles for linear, symmetric, and asymmetric top molecules. The account of molecular rotation results in rich behavior at short inter-molecular distances, with a strong dependence on both the direction of the external field and the separation. The molecules examined largely share a similar behavior, with symmetric and asymmetric molecules adding an increasing complexity, resulting in a high number of states with similar energies and dipole projections at short separations. 

We found that the dipole projection on the direction of the electric field becomes enhanced at short separation if the field is in the direction of the vector connecting the molecules, resulting in a stronger polarization. In contrast, for fields orthogonal to the molecules' plane, at short separations, it is not favorable to have dipoles polarized in the direction of the field due to the strong repulsive dipole-dipole interaction. These findings are consistent with a previous related work~\cite{lepers_long-range_2013}. While these effects arise at very short separations, these might be relevant to account for many-body applications of polar molecules when building effective dipole-dipole interactions. More generally, the account of rotation in many-body applications of polar molecules could be exploited to achieve new forms of quantum matter not realizable by single atoms, as also suggested by related few-body works on rotating diatomic molecules~\cite{dawid_two_2018,dawid_magnetic_2020}.

The rich behavior shown by the molecules, especially at short separations, could be also relevant to consider when proposing polar molecules as qubit storage~\cite{gregory_robust_2021}, as the dipole-dipole interaction plays an important role in designing the corresponding quantum gates~\cite{ni_dipolar_2018}. While diatomic molecules have attracted the most attention in the past due to their higher level of control, which is especially useful for dipolar gases, polyatomic molecules could be exploited in other applications. Qubits constructed with symmetric tops have been proposed~\cite{wei_entanglement_2011}, which rely on having opposite dipole projections for states with $k=\pm 1$. Therefore, a good understanding of the rotational properties of interacting symmetric molecules is necessary for the precise control of the dipole polarization required for designing quantum computing platforms~\cite{yu_scalable_2019}.

While the model employed in this work does not aim to describe all aspects of realistic cold molecules, it serves as a starting point for future realistic descriptions.
Future work will consider additional effects, such as from hyperfine structure and interactions with external magnetic and ac electric fields~\cite{krems_molecules_2019}. The consideration of higher-order multipolar interactions is also a pressing extension~\cite{craig_molecular_1984}, as the consideration of additional interactions can compete with the dipole-dipole one considered in this work~\cite{lepers_long-range_2013}. We also intend to consider mobile molecules trapped in harmonic traps, as studied for two molecules in Refs.~\cite{dawid_two_2018,dawid_magnetic_2020}, with the final aim of describing more realistic many-body configurations. In particular, the use of electric fields and dipole-dipole interactions could be examined to realize gases with bound tetramers of diatomic molecules. Similarly, we intend to study polar molecules immersed in lattice configurations, complementing recent efforts to describe non-reactive molecules in optical lattices~\cite{docaj_ultracold_2016,wall_microscopic_2017}. In this direction, it will be particularly interesting to examine non-linear molecules to further describe recently proposed scenarios of symmetric and asymmetric molecules trapped in lattice configurations~\cite{yu_scalable_2019,isaule_quantum_2022}.

\section*{Acknowledgements}
We acknowledge funding from EPSRC (UK) through Grant No. EP/V048449/1. J.B.G. acknowledges funding from the Leverhulme Trust.

\appendix

\section{Single-molecule physics}
\label{app:single}

In the following, we summarize the description of a single rotating polar molecule for readers unfamiliar with rotating molecules and spherical tensors. For comprehensive reviews, we refer to Refs.~\cite{zare_angular_1988,krems_molecules_2019,di_lauro_rotational_2020}.

\subsection{Rotational Hamiltonian}
\label{app:single;sub:rot}

The Hamiltonian describing a single rigid rotor reads
\begin{equation}
    \Hop_\mathrm{rot}=\left(A\JJop_{a}^2+B\JJop_{b}^2+C\JJop_{c}^2\right)\,.
    \label{app:single;sub:rot;Hrot}
\end{equation}
The rotational constants are defined from the principal moments of inertia $I_a$, $I_b$, $I_c$, as $A=\hbar/(4\pi I_a)$, $B=(\hbar/4\pi I_b)$, and $C=(\hbar/4\pi I_c)$, so they have units of frequency. The values of the rotational constants define the geometry of the molecule. Indeed, all molecules can be classified as
\begin{align*}
    I_a=0,\,I_b=I_c>0 \quad& \textrm{: Linear top} \,,\\
    I_a=I_b=I_c \quad& \textrm{: Spherical top} \,,\\
    I_a=I_b<I_c \quad& \textrm{: Oblate symmetric top} \,,\\
    I_a<I_b=I_c \quad& \textrm{: Prolate symmetric top} \,,\\
    I_a<I_b<I_c \quad& \textrm{: Asymmetric top} \,.
\end{align*}
Diatomic and some triatomic molecules are linear tops, while many small polyatomic molecules, such as CH$_3$F and CaOCH$_3$ are symmetric tops. However, most polyatomic molecules are asymmetric tops, even though some can approximately be treated as near symmetric. 

\begin{figure}[t]
\centering
\includegraphics[width=\columnwidth]{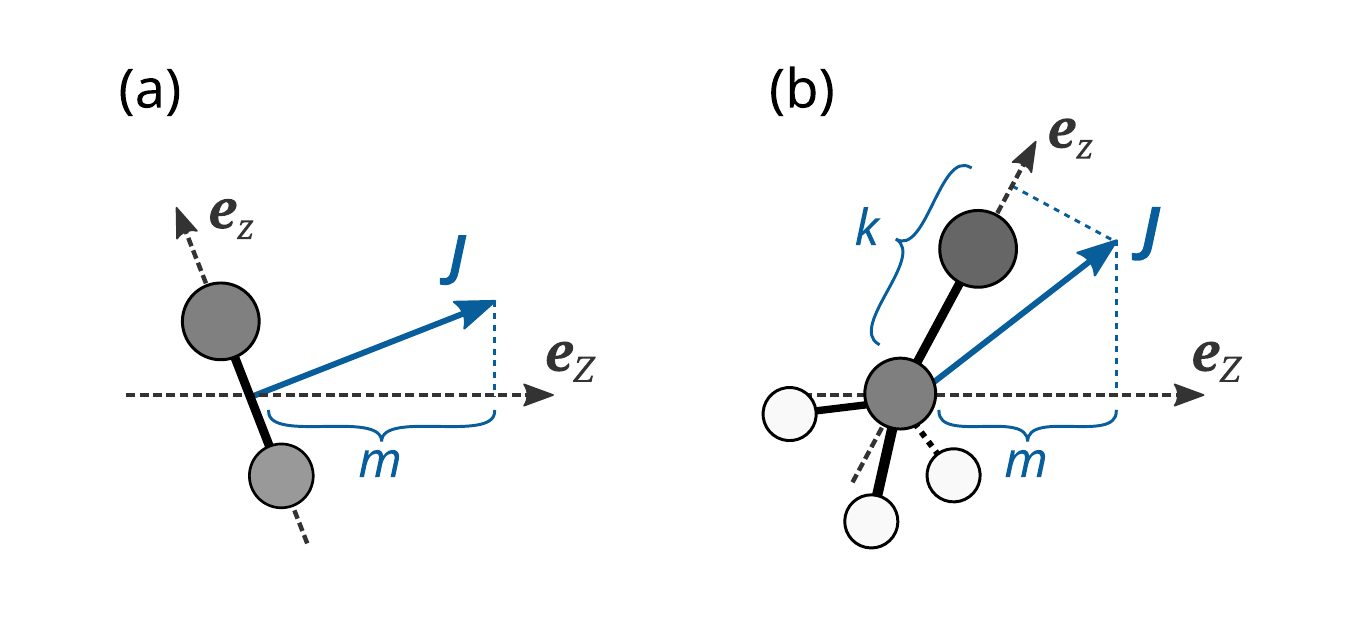}
\caption{Illustration of a rotating linear (a) and prolate symmetric (b) top molecule with angular momentum $\bm{J}$ with respect to the laboratory-fixed axis $\bm{e}_Z$ and the molecule-fixed axis $\bm{e}_z$.}
\label{app:single;sub:rot;fig:jkm}
\end{figure}

To diagonalize (\ref{app:single;sub:rot;Hrot}), one generally works in terms of the symmetric top wavefunctions $\ket{j\,k\,m}$, where $j=0,1,...$, $-j\leq k \leq j$, and $-j\leq m \leq j$. These are defined as
\begin{equation}
    \braket{\Omega }{j\,k\,m} = \sqrt{\frac{2j+1}{8\pi}}D^{j*}_{m,k}(\Omega),
\end{equation}
where here $\Omega=(\phi,\theta,\chi)$ denotes the Euler angles between the molecule- and laboratory-fixed frames, and $D^{j}_{m,k}(\Omega)$ are the Wigner $D$-matrices~\cite{zare_angular_1988}. In the $\ket{j\,k\,m}$ basis, the angular momentum operator satisfies
\begin{align}
\JJop^2 \, \ket{j\,k\,m} &= j(j+1)\,\ket{j\,k\,m}\,, \label{app:single;sub:rot;J2}\\
\Jop_z \, \ket{j\,k\,m} &= k\,\ket{j\,k\,m}\,, \label{app:single;sub:rot;Jz}\\
\Jop_Z \, \ket{j\,k\,m} &= m\,\ket{j\,k\,m}\,. \label{app:single;sub:rot;JZ}
\end{align}
 As illustrated in Fig.~\ref{app:single;sub:rot;fig:jkm}, $k$ represents the projection of the angular moment onto the molecule-fixed frame, whereas $m$ represents the usual projection onto the laboratory-fixed frame. Therefore, $\ket{j\,k\,m}$ enables one to identify the rotational state of the molecule. Note that, as explained in the main text, for symmetry arguments we define $(x, y, z) = (a, b, c)$ for oblate symmetric tops, whereas as $(x, y, z) = (b, c, a)$ otherwise~\cite{zare_angular_1988}. Also note that linear molecules cannot rotate on their $a$-axis, so we can simply consider $k=0$.

In the case of linear and spherical tops, it is easy to see that the rotating Hamiltonian is simply
\begin{equation}
    \Hop_\mathrm{rot} = B\JJop^2\,,
    \label{app:single;sub:rot;Hrot_l}
\end{equation}
resulting in the textbook spectrum $E_j=Bj(j+1)$. In the case of prolate tops, the Hamiltonian can be written as
\begin{equation}
   \Hop_\mathrm{rot} = C\JJop^2+(A-C)\Jop_z^2\,.
    \label{app:single;sub:rot;Hrot_pl}
\end{equation}
From Eqs.~(\ref{app:single;sub:rot;J2}) and (\ref{app:single;sub:rot;Jz}), one sees that $\Hop_\mathrm{rot}$ is diagonal in $\ket{j\,k\,m}$, resulting in the spectrum $E_{j,k}=Cj(j+1)+(A-C)k^2$. Similarly, for oblate tops one simply exchanges $A$ by $C$ in Eq.~(\ref{app:single;sub:rot;Hrot_pl}), resulting in $E_{j,k}=Aj(j+1)+(C-A)k^2$.

Finally, asymmetric tops do not have $\Hop_\mathrm{rot}$ diagonal  in $\ket{j\,k\,m}$. Indeed, the non-zero elements of the Hamiltonian matrix are given by
\begin{equation}
    \langle j\,k\,m |\Hop_\mathrm{rot}|j\,k\,m\rangle=\frac{B+C}{2}\left(j(j+1)-k^2\right)+A\,k^2\,,\label{app:single;sub:rot;Hrot_adiag}
\end{equation}
and
\begin{equation}
     \langle j\,k\,m |\Hop_\mathrm{rot}|j\,k\pm 2\,m\rangle=\frac{B-C}{4}f_\pm(j,k)\,,\label{app:single;sub:rot;Hrot_nonadiag}
\end{equation}
where 
\begin{align}
    f_\pm(j,k)=&\Big(\left[j(j+1)-k(k\pm 1)\right]\nonumber\\
    &\times\left[j(j+1)-(k\pm 1)(k\pm 2)\right]\Big)^{1/2}\,.
\end{align}
Therefore, the Hamiltonian couples states with $k\pm 2$, requiring a numerical diagonalization. We refer to Refs.~\cite{zare_angular_1988,koch_quantum_2019} for details on the spectrum of asymmetric tops. In addition, because $k$ is not a good quantum number, the wavefunctions of asymmetric tops take the form $\ket{j\,\tau\,m}=\sum_k c_{\tau,k} \ket{j\,k\,m}$, where $\tau=k_a-k_c=-j,...,+j$ is a pseudo-quantum number that labels the states, where $k_a=0,..,j$ and $k_c=0,...,j$ are the values that $\abs{k}$ would take in the prolate and oblate limits, respectively~\cite{zare_angular_1988}.

\subsection{Spherical tensors}
\label{app:single;sub:sph}

In order to describe interacting rotating molecules, one needs to adapt the interaction terms of the Hamiltonian to the $\ket{j\,k\,m}$ basis. This is easily done by using spherical tensors, which enable one to systematically transform vectors between the laboratory- and molecule-fixed frames.

An $\ell$th-rank spherical tensor $T$ has $2\ell+1$ elements with labels $s=-\ell,...,\ell$. Spherical tensors transform as the following~\cite{zare_angular_1988}
\begin{align}
    T_p^{(\ell)}=&\sum_{q=-\ell}^\ell D^{\ell*}_{p,q}(\Omega)T_q^{(\ell)}\,,\\
    T_q^{(\ell)}=&\sum_{p=-\ell}^\ell D^\ell_{p,q}(\Omega)T_p^{(\ell)}\,,
    \label{app:single;sub:sph;eq:T}
\end{align}
where $p$ and $q$ denote elements in the laboratory- and molecule-fixed frame, respectively.  The tensor product between two tensors $V$ and $U$ in an arbitrary frame is defined as
\begin{align}
    [V^{(\ell_v)}\otimes U^{(\ell_u)}]^{(\ell)}_s=&\sum_{s_v,s_u} \braket{\ell_v\,s_v, \ell_u\,s_u}{\ell\,s} V^{(\ell_v)}_{s_v}U^{(\ell_u)}_{s_u}\nonumber\\
    =&\sum_{s'=-\ell}^\ell \braket{\ell_v\,s', \ell_u\,s-s'}{\ell\,s}  V^{(\ell_v)}_{s'} U^{(\ell_u)}_{s-s'}\,,
    \label{app:single;sub:sph;eq:Tprod}
\end{align}
where $\braket{l_1\,m_1,l_2\,m_2}{L\,M}$ are Clebsch–Gordan coefficients, and $\ell_v$ and $\ell_u$ are the ranks of tensors $V$ and $U$, respectively. 

In this work, we transform vectors to rank-1 spherical tensors by employing the spherical basis $(\bm{e}_{-1},\bm{e}_{0},\bm{e}_{1})$. In an arbitrary frame with Cartesian coordinates $(x',y',z')$, the unit spherical vectors read
\begin{equation}
    \bm{e}_0 = \bm{e}_{z'},\qquad \bm{e}_{\pm 1}=\mp(\bm{e}_{x'}\pm i\,\bm{e}_{y'})/\sqrt{2}\,.
    \label{app:single;sub:sph;eq:e}
\end{equation}
The $s=-1,0,+1$ spherical elements of a vector $\bm{v}=(v_{x'},v_{y'},v_{z'})$ in this frame are given by $v_s=\bm{e}_s\cdot\bm{v}=v C^{(1)}_s(\Omega_{\bm{v}})$, where 
\begin{equation}
    C^{(l)}_m(\Omega)=D^{l*}_{m,0}(\Omega)=\sqrt{\frac{4\pi}{2l+1}}Y_{l,m}(\Omega)\,,
    \label{app:single;sub:sph;eq:C}
\end{equation}
are the unnormalized spherical harmonics, and $\Omega_{\bm{v}}=(\theta_{\bm{v}},\phi_{\bm{v}})$ denotes the polar and azimuthal angles of $\bm{v}$ in the chosen frame. Therefore, the spherical components of $\bm{v}$ can be expressed as
\begin{equation}
    v_0 = v_{z'},\qquad v_{\pm 1}=\mp(v_{x'}\pm i\,v_{y'})/\sqrt{2}\,.
    \label{app:single;sub:sph;eq:v}
\end{equation}
All the usual vector operations can be defined in terms of spherical tensors~\cite{zare_angular_1988}. However, in this work, we are only interested in the dot product between two vectors. This takes the simple form
\begin{equation}
    \bm{v}^*\cdot\bm{u}=\sum_{s=-1}^{1}(-1)^s v_s u_{-s}\,.
    \label{app:single;sub:sph;eq:dot}
\end{equation}

The previous equations enable one to perform all the necessary transformations between frames. Indeed, the rotational state is encoded entirely in the $D$-matrices, which appear from the transformations~(\ref{app:single;sub:sph;eq:T}). The matrix elements of the Hamiltonian in the $|j\,k\,m\rangle$ basis are then obtained by using that~\cite{zare_angular_1988}
\begin{align}
    \mel{j\,k\,m}{D^{J*}_{M,K}}{j'\,k'\,m'} =&(-1)^{m+k}\sqrt{(2j+1)(2j'+1)}\nonumber\\
    &\times\threej{j}{J}{j'}{-m}{M}{m'}\threej{j}{J}{j'}{-k}{K}{k'}\,\,
    \label{app:single;sub:sph;eq:DcJKM}
\end{align}
where the parentheses are the Wigner 3j-symbols. Note that the only non-zero matrix elements are those with $m=m'+M$ and $k=k'+K$.

\subsection{Interacting molecule: Stark effect}
\label{app:single;sub:dc}

\begin{figure*}[t]
\centering
\includegraphics[width=\textwidth]{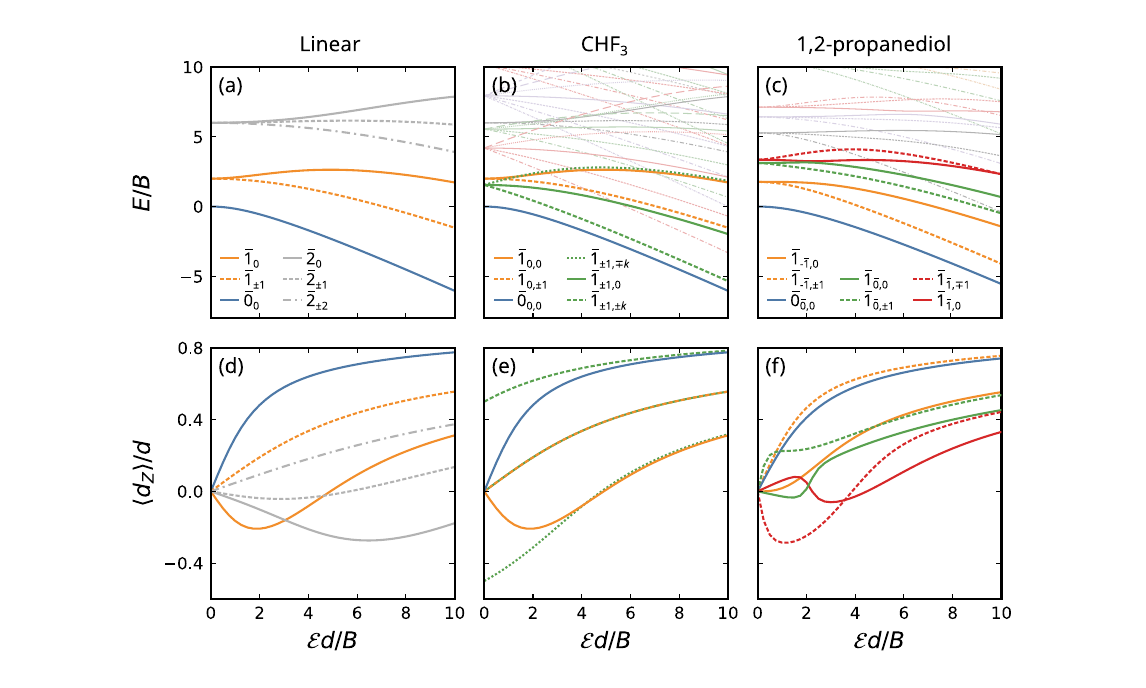}
\caption{Energy spectrum (a,b,c) and average projection of the dipole moment onto the $\bm{e}_Z$ axis (d,e,f) as a function of the magnitude of the external electric field. In the figure $d=|\bm{d}|$ and $B$ is the rotational constant in the $b$-axis. Panels (a,d) show results for a linear molecule, (b,e) show results for CHF$_3$, and (c,f) show results for 1,2-propanediol.}
\label{app:single;sub:dc;fig:E_dZ}
\end{figure*}

We now employ the previous definitions to describe a single rotating molecule interacting with an external dc electric field. The system is described by
\begin{equation}
    \Hop = \Hop_\mathrm{rot}+\Hop_\mathrm{dc}\,,
\end{equation}
where $\Hop_\mathrm{rot}$ is given by Eq.~(\ref{app:single;sub:rot;Hrot}) and $\Hop_\mathrm{dc}=-\bm{d}\cdot \bm{\mathcal{E}}$, with $\bm{d}$ the permanent dipole of the molecule and $\bm{\mathcal{E}}$ the chosen electric field. The matrix elements of $\Hop_\mathrm{rot}$ in $\ket{j\,k\,m}$ are given in~\ref{app:single;sub:rot}. To compute the matrix elements of $\Hop_\mathrm{dc}$, both $\bm{d}$ and $\bm{\mathcal{E}}$ are written as spherical tensors, resulting in
\begin{align}
   \Hop_\mathrm{dc}&=-\sum_{p=-1}^1(-1)^p d_p \mathcal{E}_{-p}\,,\\
    &=-\sum_{p,q}(-1)^pD^{1*}_{p,q}(\Omega)\,d_q \mathcal{E}_{-p}\,. \label{app:single;sub:dc;eq:Hdcpq}
\end{align}
where in the first line we used Eq.~(\ref{app:single;sub:sph;eq:dot}) with both tensors in the laboratory-fixed frame, and in the second line we transformed $\bm{d}$ to the molecule-fixed frame using Eq.~(\ref{app:single;sub:sph;eq:T}). Here $d_q$ and $\mathcal{E}_p$ are the inputs of the calculation. The matrix elements of Eq.~(\ref{app:single;sub:dc;eq:Hdcpq}) then read
\begin{align}
\mel{j\,k\,m}{\Hop_\mathrm{dc}}{j'\,k'\,m'} =&-\sum_{p,q}(-1)^p\,d_q \mathcal{E}_{-p}\nonumber\\
&\times\mel{j\,k\,m}{D^{1*}_{p,q}(\Omega)}{j'\,k'\,m'}\,,
\end{align}
where $\mel{j\,k\,m}{D^{1*}_{p,q}}{j'\,k'\,m'}$ is given by Eq.~(\ref{app:single;sub:sph;eq:DcJKM}).

The energy spectrum is obtained by diagonalizing $\Hop$ either numerically, as in this work, or perturbatively for weak fields~\cite{krems_molecules_2019}. In addition, the average value of the dipole moment projection in some laboratory-frame direction $p$ is computed from $\expval{d_p}=\expval{d_p}{\Psi}$, where $\ket{\Psi}=\sum_{\alpha} c_{\alpha} \ket{j_\alpha k_\alpha m_\alpha}$ is a chosen eigenvector solution. Therefore,
\begin{equation}
    \expval{d_p}{\Psi} = \sum_{\alpha,\beta} c^*_{\alpha}c_{\beta}\sum_{q} d_q 
    \mel{j_\alpha k_\alpha m_\alpha}{D^{1*}_{p,q}(\Omega)}{j_\beta k_\beta m_\beta}\,,
\end{equation}
where $d_q$ are the known elements of the dipole moment in the molecule-fixed frame. After computing $\expval{d_p}$, one can extract the projections in cartesian coordinates $(X,Y,Z)$ by using Eq.~(\ref{app:single;sub:sph;eq:v}).

We show textbook examples~\cite{krems_molecules_2019,wall_simulating_2013} of the low-energy spectrum and dipole projection as a function of the external dc field in Fig.~\ref{app:single;sub:dc;fig:E_dZ}. Only the lower-energy states are highlighted for readability. As in the main text, we show examples for linear molecules, the oblate symmetric top CHF$_3$, and 1,2-propanediol. We choose a dc field $\bm{\mathcal{E}}=\mathcal{E}\bm{e}_Z$, and so the projection of the dipole moment is only finite on the $Z$-axis. However, for a single rotating molecule, the chosen direction of $\bm{\mathcal{E}}$ is arbitrary.

We label the states as $\bar{j}_m$ for linear molecules (a), as $\bar{j}_{k,m}$ for symmetric molecules (b), and as $\bar{j}_{\bar{\tau},m}$ for asymmetric molecules (c), where $j$ is the eigenvalue of $\bm{J}^2$ for $\mathcal{E}=0$ and $\tau$ is the pseudo-quantum number relevant to asymmetric tops. Note that while $m$ remains conserved for finite electric fields, $\Hop_\mathrm{dc}$ couples states with different $j$~\cite{krems_molecules_2019}. 

For $\mathcal{E}=0$, the spectrum is that of a non-interacting rotating molecule described in \ref{app:single;sub:rot}. With a finite $\mathcal{E}$, the spectrum shows the usual Stark effect, with a splitting of the rotational states. Naturally, symmetric and asymmetric tops show a much richer spectrum due to the additional projection $k$. The lines with finite values of $k$ (or $\tau$) and $m$ are twice degenerate, with solutions where $k$ and $m$ having equal signs (denoted by $\pm m$) and solutions where $k$ and $m$ having opposite signs (denoted by $\mp m$). 

The dipole moment's projection naturally increases with $\mathcal{E}$, aligning the molecule in the direction of the field. However, $\expval{d_Z}$ still depends strongly on the internal state of the molecule. Therefore, one can only neglect the rotation of the molecules for strong fields $\mathcal{E}d\gg B$~\cite{lahaye_physics_2009}.

We refer to Refs.~\cite{krems_molecules_2019,di_lauro_rotational_2020} for more detailed analyses of molecules interacting with electric fields. We also refer to Ref.~\cite{strebel_improved_2010} for a detailed analysis of the rotational spectrum of CHF$_3$.

\begin{widetext}
\section{Matrix elements}
\label{app:matrix}

In the following we provide the matrix elements of $\Hop$ [Eq.~(\ref{sec:model;sub:H:eq:H})] in the basis $\ket{j_1\,k_1\,m_1,j_2\,k_2\,m_2}$. For compactness, we use the notation $\mel{1,2}{\Hop}{1',2'}=\mel{j_1\,k_1\,m_1,j_2\,k_2\,m_2}{\Hop}{j'_1\,k'_1\,m'_1,j'_2\,k'_2\,m'_2}$.

The rotational and dc terms of the Hamiltonian are simply given by the corresponding terms for the individual molecules
\begin{align}
    \mel{1,2}{\Hop_\mathrm{rot/dc}}{1',2'}=&\mel{j_1\,k_1\,m_1}{\Hop_{\mathrm{rot/dc},1}}{j'_1\,k'_1\,m'_1}\delta_{j_2,j'_2}\delta_{k_2,k'_2}\delta_{m_2,m'_2}\nonumber\\
    &+\mel{j_2\,k_2\,m_2}{\Hop_{\mathrm{rot/dc},2}}{j'_2\,k'_2\,m'_2}\delta_{j_1,j'_1}\delta_{k_1,k'_1}\delta_{m_1,m'_1}\,,
\end{align}
where $\mel{j_i\,k_i\,m_i}{\Hop_{\mathrm{rot},i}}{j'_i\,k'_i\,m'_i}$ is given by the matrix elements described in Appendix~\ref{app:single;sub:rot} [Eqs.~(\ref{app:single;sub:rot;Hrot_l}--\ref{app:single;sub:rot;Hrot_nonadiag})], whereas $\mel{j_i\,k_i\,m_i}{\Hop_{\mathrm{dc},i}}{j'_i\,k'_i\,m'_i}$ is given by Eq.~(\ref{app:single;sub:dc;eq:Hdcpq}).

To compute the matrix elements for $\Hop_\mathrm{dd}$ we start from the simplified form for our chosen laboratory frame~(\ref{sec:model;sub:H;eq:HddsphZ}). Because the dipole moments in Eq.~(\ref{sec:model;sub:H;eq:HddsphZ}) are expressed in the laboratory-fixed frame, we transform them to their known molecule-fixed values using~(\ref{app:single;sub:sph;eq:T}). $\Hop_\mathrm{dd}$ then reads:
\begin{equation}
    \Hop_\mathrm{dd} = -\frac{1}{r^3}\sum_{q_1,q_2}\left[2D^{1*}_{0,q_1}(\Omega_1)D^{1*}_{0,q_2}(\Omega_2)+D^{1*}_{-1,q_1}(\Omega_1)D^{1*}_{1,q_2}(\Omega_2)+D^{1*}_{1,q_1}(\Omega_1)D^{1*}_{-1,q_2}(\Omega_2)\right]d_{1,q_1}d_{2,q_2}\,,
\end{equation}
where $q_1$ and $q_2$ run over each molecule-fixed frame. Therefore, the matrix elements are
\begin{align}
    \mel{1,2}{\Hop_\mathrm{dd}}{1',2'}=&-\frac{1}{r^3}\Big[2\mel{j_1\,k_1\,m_1}{D^{1*}_{0,k_1-k'_1}}{j'_1\,k'_1\,m'_1}\mel{j_2\,k_2\,m_2}{D^{1*}_{0,k_2-k'_2}}{j'_2\,k'_2\,m'_2}\delta_{0,m_1-m'_1}\delta_{0,m_2-m'_2}\nonumber\\
    &+\mel{j_1\,k_1\,m_1}{D^{1*}_{-1,k_1-k'_1}}{j'_1\,k'_1\,m'_1}\mel{j_2\,k_2\,m_2}{D^{1*}_{1,k_2-k'_2}}{j'_2\,k'_2\,m'_2}\delta_{-1,m_1-m'_1}\delta_{1,m_2-m'_2}\nonumber\\
    &+\mel{j_1\,k_1\,m_1}{D^{1*}_{1,k_1-k'_1}}{j'_1\,k'_1\,m'_1}\mel{j_2\,k_2\,m_2}{D^{1*}_{-1,k_2-k'_2}}{j'_2\,k'_2\,m'_2}\delta_{1,m_1-m'_1}\delta_{-1,m_2-m'_2}\Big]\nonumber\\
    &\times d_{1,k_1-k'_1}d_{2,k_2-k'_2} \,,
\end{align}
where the matrix elements of the $D$-matrices are given by Eq.~(\ref{app:single;sub:sph;eq:DcJKM}). Importantly, most of the matrix elements of all terms in the Hamiltonian are zero. This enables us to work with sparse matrices in our numerical analysis, which is essential to work with the truncated basis used in this work. Otherwise, we would only be able to work up to a cutoff of $j_\mathrm{max}=4$ or less.

\begin{figure*}[t]
\centering
\includegraphics[width=\textwidth]{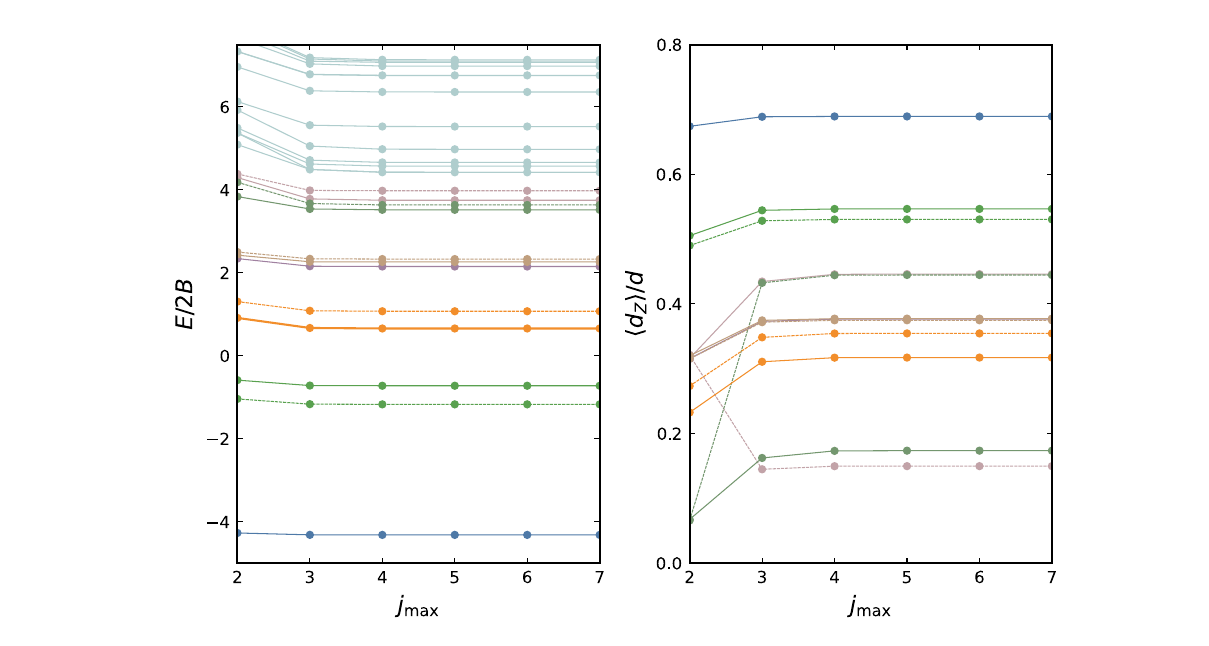}
\caption{Potential-energy curves and average dipole moment projections $\expval{d_Z}$ for linear molecules as a function of the angular momentum cutoff $j_\mathrm{max}$. We consider a distance $r=r_B$ and an external field $\bm{\mathcal{E}}d/B=4\bm{e}_Z$. Lines with the same color and line type in the left and right panels indicate the same eigenstate.}
\label{app:convergence:fig:linear}
\end{figure*}

\end{widetext}

\section{Convergence}
\label{app:convergence}

To illustrate the dependence of our numerical calculations on the choice of $j_\mathrm{max}$, in Fig.~\ref{app:convergence:fig:linear} we show examples of the potential-energy curves and dipole projection as a function of this cutoff. The figure shows results for linear molecules and a particular choice of separation and field, but similar results are obtained for other choices.

\begin{figure*}[t]
\centering
\includegraphics[width=\textwidth]{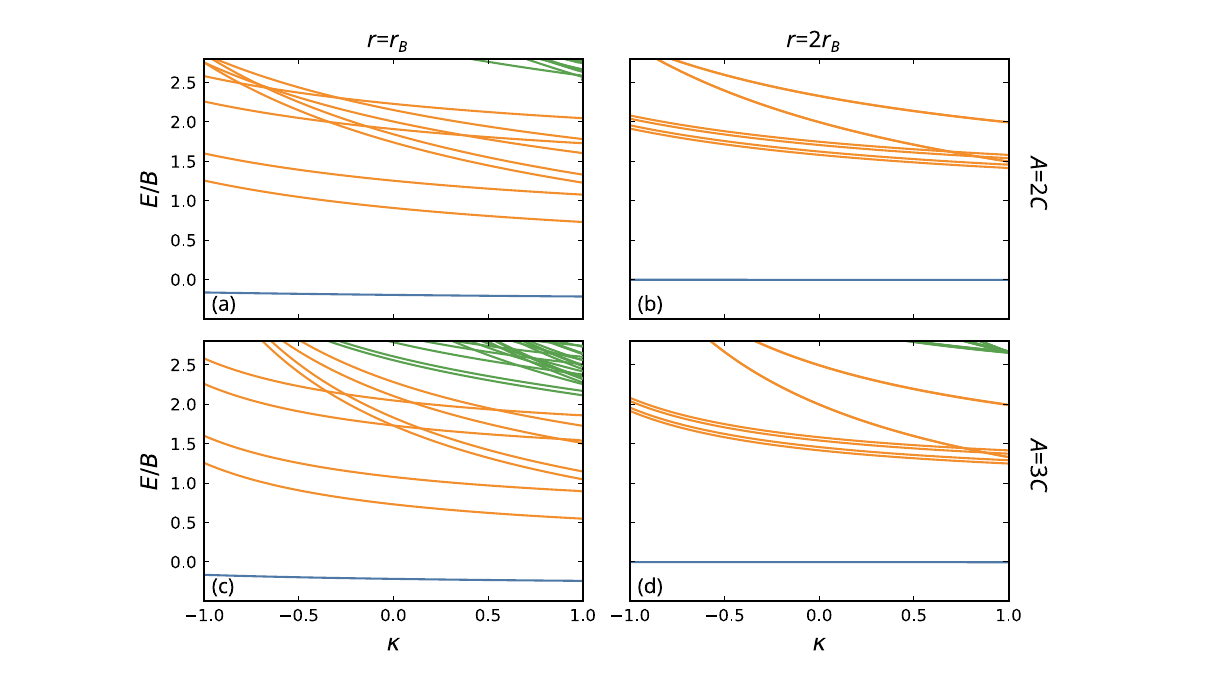}
\caption{Potential-energy curves for two asymmetric top molecules as a function of the asymmetry parameter $\kappa$. We consider a distance $r=r_B$ (left panels) and $r=2r_B$ (right panels), and no external electric field. We consider $A=2C$ (top panels) and $A=3C$ (bottom panels). Blue lines correspond to states with $\bar{j}=0$, orange lines correspond to states with $\bar{j}=1$, and green lines correspond to states with $\bar{j}=2$.}
\label{app:asym:fig:E0}
\end{figure*}

For the low potential-energy curves examined in this work, the figure shows that the results quickly converge for $j_\mathrm{max}\gtrsim 4$, even though the dipole projection seems to be more sensitive to $j_\mathrm{max}$ than the energies. Because the dipole-dipole interaction and weak external fields only couple Fock states with small differences in $j$,  a cutoff of $j_\mathrm{max}= 4$ already captures the relevant contributions for the states with $\bar{j}\lesssim 2$ examined in this work. In addition, the figure shows clearly that the quality of the results decreases for small $j_\mathrm{max}$ and for higher excited states (see for example the light cyan lines in the potential-energy curves). Naturally, to access higher parts of the energy spectrum it is necessary to consider bases with a higher cutoff.

Finally, we mention that a similar convergence with $j_\mathrm{max}$ is obtained for the low-energy curves with other rotational constants, separations, and external field strengths. However, in the diverging region of the potential-energy curves $r\ll r_B$, we show many higher-energy curves [see for example the non-highlighted curves in Fig.~\ref{sec:noE;fig:spectrum}]. Because some of these higher-energy curves have much higher values of $j$, those curves may indeed not converge completely for $j_\mathrm{max}=7$. Nevertheless, those higher-energy curves will still diverge at short distances and their examination is beyond the scope of this work.

\section{Asymmetric molecules}
\label{app:asym}

To examine how the eigenstates of two asymmetric top molecules depend on the values of the rotational constants, in Fig.~\ref{app:asym:fig:E0} we show the potential-energy curves at a fixed rescaled distance $r/r_B$ and no external field $\mathcal{E}=0$ as a function of the asymmetry parameter~\cite{zare_angular_1988}
\begin{equation}
    \kappa = \frac{2B-A-C}{A-C}\,,
\end{equation}
where $\kappa=-1$ in the limit of prolate symmetric tops and $\kappa=1$ in the limit of oblate symmetric tops. he values of $A$ and $C$ remain fixed, and thus only $B$ changes with $\kappa$. The color scheme employed only differentiates states with different $\bar{j}$, as the labeling system used in the main text cannot clearly connect the prolate and oblate limits~\cite{zare_angular_1988}.

As stressed in the main text, the potential-energy curves depend on both the rotational constants and the inter-molecular separation. For $r=2r_B$, the dipole-dipole interaction effect is weaker, and thus the potential-energy curves as a function of $\kappa$ are similar to the energy levels of single rotating asymmetric molecules (see~\cite{zare_angular_1988}). On the other hand, for $r=r_B$ the molecules exhibit a rich behavior of the potential-energy curves. As discussed, measuring the energy curves of interacting molecules could be employed to extract the moments of inertia. We intend to explore these ideas further in future work.

\section{Dependence on the electric field strength}
\label{app:E}

Sec.~\ref{sec:E} considers a finite electric field with strength $\mathcal{E}d/B=4$. To provide an example of how the results change with other choices of electric fields, we show the potential-energy curves and dipole projections as a function of the $\mathcal{E}$ for two linear molecules separated by a distance $r=r_B$. As in the main text, the figure considers the limits when the electric field is oriented in the $Z$-direction (left panels) and the $X$-direction (right panels).

\begin{figure*}[t]
\centering
\includegraphics[width=\textwidth]{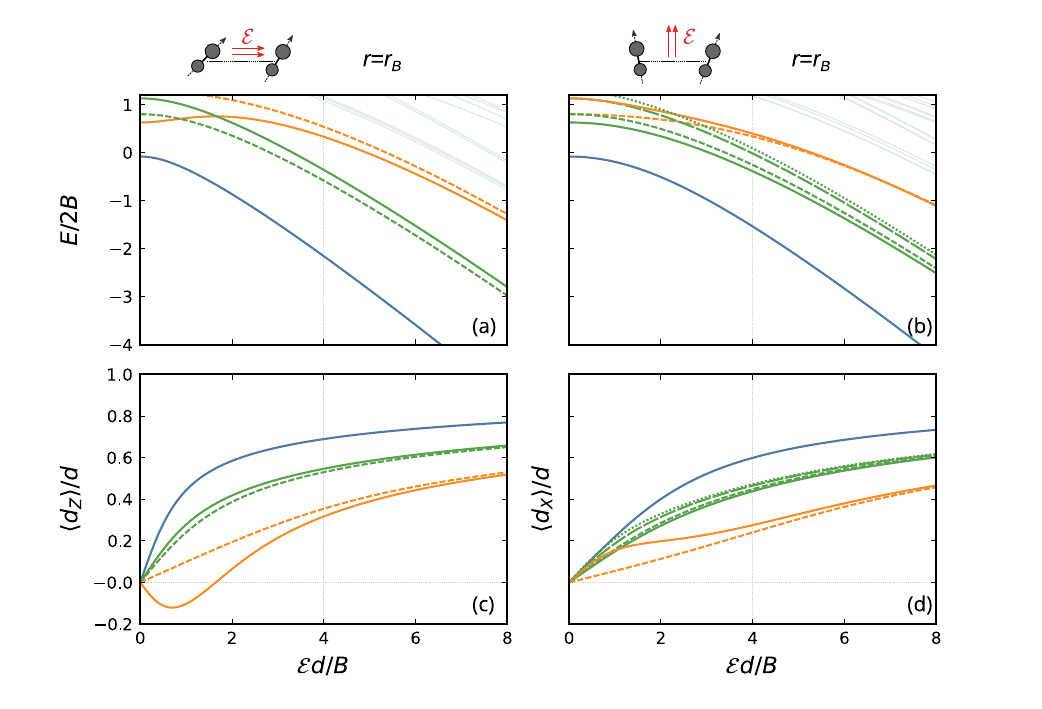}
\caption{Potential-energy curves (a,b) and average dipole moment projection on the $\bm{e}_Z$ (c) and $\bm{e}_X$ (d) axes of two linear molecules as a function of the external electric field strength $\mathcal{E}$. The left panels (a,c) consider an external dc field in the $X$-direction $\bm{\mathcal{E}}=\mathcal{E}\bm{e}_Z$, whereas the right panels (b,d) consider an external dc field in the $X$-direction $\bm{\mathcal{E}}=\mathcal{E}\bm{e}_X$. Lines with the same color and line type in the upper and bottom panels indicate the same eigenstate.  The lines and colors denote the same convention used in Fig.~\ref{sec:E;sub:linear;fig:r}.}
\label{app:E;fig:E}
\end{figure*}

From the top panels, we observe that while for weak electric fields, some levels can increase their energy, for larger $\mathcal{E}$ the energy decreases. The latter can be expected from the known behavior of single molecules [see top panels of Fig.~\ref{app:single;sub:dc;fig:E_dZ}]. More importantly,  the energy curves show crossings when $\mathcal{E}$ and $B/d$ are comparable. By noting that the figure also considers $r=r_B$, the crossings are a result of the competition between the electric field, rotation, and the dipole-dipole interaction. For stronger electric fields the energy curves decrease monotonously without showing crossings. Therefore, to achieve molecular rotational control, considering the regime $\mathcal{E}\gg B/d$ could ease the separation of the energy levels.

Finally, the bottom panels show the gradual polarization of the molecules from zero to the direction of the electric field. Once again, the panels show a richer behavior for weak electric fields, even showing a negative polarization. The latter is also expected from the single-molecule behavior [see bottom panels of Fig.~\ref{app:single;sub:dc;fig:E_dZ}]. Nevertheless, for stronger electric fields all the dipole projections become positive and increase monotonously, as expected.

We note that similar behavior is observed for other molecules.

\newpage

\section*{References}
\bibliography{biblio}

\end{document}